\documentclass{aa}

\usepackage{graphicx}
\usepackage{amsmath}
\usepackage{txfonts}
\usepackage{lipsum}
\usepackage{subcaption}         % necessary for continued figures, example in section 3
\usepackage{lscape}             % to rotate a single page table, example in appendix.
\usepackage{placeins}           % useful with \FloatBarrier, to keep 
\usepackage{hyperref}
\hypersetup{colorlinks=true,linkcolor=blue,citecolor=blue,urlcolor=blue}
\newcommand{\xrism}{XRISM}

\newcommand{\rev}[1]{#1}

\begin{document}

   %\title{An Analytic Approach to Accretion Disc Wind Emission Lines}
    \title{Unveiling BLR Structure in AGN with High Resolution X-ray Spectra}

   %\subtitle{Models for the era of high resolution X-ray spectroscopy}
   \subtitle{An Analytic Approach to Wind Emission Line Profiles}

%%%%%%%%%%%%%%%%%%%%%%%%%%%%%%%%%%%%%%%%
% Please do not include ORCIDs next to author names.
% Only ORCIDs authenticated by individual authors in EDP Sciences editorial system will be taken into account.
% ORCIDs included here will be removed.
%%%%%%%%%%%%%%%%%%%%%%%%%%%%%%%%%%%%%%%%

   \author{Scott Hagen\inst{1,2,3,4}
        \and Chris Done\inst{4} \and Gabriele A. Matzeu\inst{5} \and Hirofumi Noda\inst{6}
        }

   \institute{
   IFPU - Institute for Fundamental Physics of the Universe, Via Beirut 2, 34151 Trieste, Italy\\
    \email{shagen@sissa.it}
    \and SISSA - International School for Advanced Studies, Via Bonomea 265, 34136 Trieste, Italy
    \and INAF - Osservatorio Astronomico di Trieste, Via G. B. Tiepolo 11, I-34143 Trieste, Italy
    \and Centre for Extragalactic Astronomy, Department of Physics, Durham University, South Road, Durham DH1 3LE, UK
    \and Quasar Science Resources SL for ESA, European Space Astronomy Centre (ESAC), Science Operations Department, 28692, Villanueva de la Ca\~{n}ada, Madrid, Spain
    \and Astronomical Institute, Tohoku University, 6-3 Aramakiazaaoba, Aoba-ku, Sendai, Miyagi 980-8578, Japan
    }

   \date{Received September 30, 20XX}

% \abstract{}{}{}{}{}
% 5 {} token are mandatory
 
  \abstract
    {
    \xrism\ has provided an unprecedented view of the emission and absorption lines in the X-ray. Notably, early results showed significant complexity to the Fe-K$\alpha$ line profile in AGN, with clear contributions from at least three emitting structures: an inner disc, intermediary broad line region (BLR) scale material, and an outer torus. This poses a new challenge for the modelling of the emission lines, as while fast sophisticated models exist for disc line-profiles, large scale-height material is typically much more complex. In this paper we aim to address this gap, by building a fully analytic model for the emission line profiles from a wind, aimed towards BLR scale material, motivated on previous reverberation studies suggesting a wind on the inner edge of the BLR. Our approach give a physically motivated, yet computationally fast, model for the intermediary component to the Fe-K$\alpha$ complex seen in the \xrism\ data. We demonstrate our model on the \xrism\ observations of NGC\,4151 from the performance verification phase, showing that it gives a good description of the data, with physically reasonable parameters for BLR scale material. We also show that our model naturally gives the `smooth' line profile seen in the data, due to the large spatial extent of a wind. Finally, we make our model code public to the community, and name it {\sc xwind}.}
   %{Optional, leave empty if necessary.  The heading “Context” is used when needed to
%ive background information on the research conducted in the paper}
   %{Mandatory. The objectives of the paper are defined here.} 
   %{Mandatory. The methods of the investigation are outlined here}
   %{Mandatory. The results are summarized here.}
   %{Optional, leave empty if necessary.  “Conclusions” can be used to
%explicit the general conclusions that can be drawn from the paper.}
 
   \keywords{Accretion, accretion disks --
                Line: profiles --
                X-rays: galaxies
               }

    %\titlerunning{BLR structure with high resolution X-ray spectra}
    \authorrunning{S. Hagen et al.}

   \maketitle

\nolinenumbers
%%%%%%%%%%%%%%%%%%%%%%%%%%%%%%%%%%%%%%%%%%%%%%%%%%%%%%%%%%%%%%
\section{Introduction}

Accreting systems, typically understood in the context of the \citet{Shakura73} disc model, display near ubiquitous out-flowing winds. From accreting white-dwarves, to black hole binaries, and up to active galactic nuclei, outflowing material seems to be a direct consequence of accretion; either driven via magnetic fields \citep[e.g][]{BlandfordPayne82, Fukumura10, Fukumura17}, thermally through inverse Compton scattering \citep[e.g][]{Begelman83a, Begelman83b}, or radiatively via the absorption of an incident photon field \citep[e.g][]{Proga00, Proga04, King10}.

%Realistic calculations of accretion disc winds are complex, typically requiring hydrodynamical simulations \citep[e.g][]{Proga00, Higginbottom18, Tomaru19}. These are naturally expensive to run, and additionally require radiative transfer calculations for the extraction of observable spectra \citep[e.g][]{Long02, Odaka11, Matthews25}.

In AGN the picture is rather complex. In the X-ray there are the ultra-fast outflows (UFOs), winds reaching a significant fraction of the speed of light ($\sim 0.2-0.3c$) typically associated with highly accreting AGN \citep[e.g][]{Reeves09, Tombesi10, Nardini15, Matzeu17, Igo20, XRISM25_PDS456, Noda25}. In the optical/UV powerful outflows can manifest in both characteristic broad-absorption lines \citep[e.g][]{Weymann91, Leighly09, Leighly18, Choi20, Choi22} or skewed and blue-shifted emission lines. Importantly for our understanding of AGN feedback, these powerful outflows are detected across cosmic time \citep[e.g][]{Maiolino12, Bischetti17, Bischetti19, Bischetti23, Fiore17}, with a clear link between the outflow power and the accretion power in the AGN \citep[e.g][]{Feruglio15, Fiore17}.

However, the majority of AGN, especially in the local Universe, are not accreting close to the Eddington limit, and so one does not necessarily expect to see these powerful outflows. Yet, even in more typical moderately accreting AGN, there is evidence for winds; albeit with significantly less power than the more extreme UFOs. The clearest examples coming from the intensive monitoring campaigns, in particular the STORM campaigns \citep{DeRosa15}, which showed clear evidence of variable absorption, now interpreted as a disc wind on the inner edge of the BLR \citep{Dehghanian19a, Dehghanian19b, Kara21}. More recently, BLR scale outflows have also been suggested in spatially resolved Gravity/Gravity+ observations \citep[e.g.][]{Gravity24, Gravity+25}. Given the prevalence of more moderately accreting AGN and the near ubiquitous presence of the BLR, understanding the structure of these BLR scale outflows and their link to the accretion flow is critical for our overall understanding of AGN structure and evolution.

Large scale-height material has the advantage that it subtends a large solid angle as seen from the central source. Given the ubiquity of high energy X-rays in AGN \citep{Elvis94, Lusso16}, this large scale height material should see, and absorb, a significant number of X-ray photons. A fraction of these will then be re-emitted, predominantly in Fe-K$\alpha$ and Fe-K$\beta$ \citep[e.g][]{Murphy09, Tzanavaris23}, with the motion of the material imprinted onto the observed line shape. Indeed prior work using Chandra, XMM, and NuSTAR spectra have shown clearly contributions to the Fe-K$\alpha$ line from BLR scale material \citep[e.g][]{Yaqoob04, Bianchi08, Miller18, Andonie22a, Andonie22b}. This is further supported by \citet{Noda23} who showed the Fe-K$\alpha$ lagging the continuum through a changing-state event with time-scales corresponding to BLR distances.

More recently, \xrism\ observations of bright, nearby, AGN have clearly distinguished multiple components to the Fe-K$\alpha$ complex, consistently showing an intermediary component consistent with BLR scale material \citep{XRISM24_NGC4151, Bogensberger25, Miller25, Mehdipour25, Kammoun25}. This provides an opportunity, as resolving individual components to the emission lines allows for the detailed modelling of each of these emitting components. However, herein lies the challenge, as detailed modelling of wind emission lines is generally complex. Firstly, physically realistic modelling of the wind structures requires expensive hydrodynamical simulations \citep[e.g][]{Proga00, Nomura16, Nomura17, Higginbottom18}. 
%Secondly, extracting observed spectra from these simulations requires the use of radiative transfer calculations, which are often complex; although a number of sophisticated methods do currently exist \citep[e.g][]{Long02, Odaka11, Mehdipour16b, Ferland17, VanderMeulen23, Luminari23, Matthews25}. These combine to give computational bottlenecks when fitting data, especially for large response files as with \xrism. This has often led to the previous \xrism\ works using either simple Gaussian smoothing models or more sophisticated accretion disc line profile models \citep[from e.g][]{Fabian89, Dovciak04, Dauser14} to fit the intermediary BLR component. While this has the advantage that these models are fast to compute, and give an estimate of the average velocity width, they do not give information on wind specific properties (e.g mass-outflow rate, covering fraction, etc). 
This can be circumvented to some extent by implementing parametric models for the wind structure \citep[e.g][]{Knigge95, Sim08, Hagino15, Matzeu22, Luminari18, Luminari24}. which allow for rapid calculations of both the velocity and density profiles. This approach is used in a number of current AGN wind models (e.g the {\sc wine} model of \citealt{Luminari24} and the {\sc xrade} model from \citealt{Matzeu22}, among others). However, in order to extract observable spectra, radiative transfer calculations are required, regardless of whether the wind structure is calculated via simulation or parametrically. Radiative transfer calculations are typically complex, leading to computational bottlenecks; although we note a number of sophisticated methods do currently exist \citep[e.g][]{Long02, Odaka11, Mehdipour16b, Ferland17, VanderMeulen23, Luminari23, Matthews25}. Often, to speed up the fitting process, studies will then \rev{resort} to calculating model grids to create interpolation tables \citep[e.g][]{Luminari24}, or for more advanced usage train a convolutional neural net \citep{Matzeu22} (emulation). These are highly advanced models, with the ability to fit a full broad-band X-ray spectrum. However, for a slow wind on the inner edge of the BLR, the main constraining power in the X-ray should originate in the Fe-K$\alpha$ complex, and so performing full radiative transfer is likely unnecessary.

%One way of reducing the computational cost is by implementing parametric models for the wind structure \citep[e.g][]{Knigge95, Sim08, Hagino15, Matzeu22, Luminari18, Luminari24, Matthews2}, allowing for rapid calculations of both the velocity and density profile. This approach is used in a number of current AGN wind models (e.g the {\sc wine} model of \citealt{Luminari24} and {\sc xrade} model from \citealt{Matzeu22}, among others). However, these are generally tailored more towards the UFO type winds seen in the more extreme objects. Additionally, they generally focus on the full X-ray spectrum (both absorption and emission), and so require detailed radiative transfer calculations. This is slow, and so these often resort to creating interpolation tables \citep[e.g][]{Luminari24} or train convolutional neural networks \citep{Matzeu22} in order to simplify the fitting process. For a BLR like wind the main constraining power should originate in the Fe-K$\alpha$ emission line complex, and so performing full radiative transfer is likely unnecessary. 

Instead, it should be possible to perform a simplified analytic calculation. This is the main aim of this paper. Here we parametrise the wind in terms of its geometry, velocity profile, and mass-outflow rate. Through mass-conservation we then calculate the density profile. Under the assumption that the material is mostly neutral, and illuminated by a simple power-law X-ray spectrum, we perform a simplified calculation of the absorption and transmittance of the incident X-ray spectrum through the wind. This approach is analytic, and allows us to extract the number of photons re-emitted in Fe-K$\alpha$, in essence the emissivity profile. Accounting for the relativistic Doppler shift induced by the wind motion, and integrating over the emissivity profile, then gives the observed frame emission line profile. This not only provides a rapid calculation of the emission line profile, it also self-consistently calculates the equivalent width of the line. These combine to give constraining power on not only the distance from the black hole of the wind, but also the mass-outflow rate and covering fraction.

This paper is organised as follows. In section\,\ref{sec:xwind} we define the model, and give a detailed description of the calculations. Then in section\,\ref{sec:mod_properties} we explore the properties of the model and its observables. In section\,\ref{sec:application_xrism} we apply our model to a \xrism\ observation of NGC\,4151, demonstrating its ability to constrain the properties of a BLR like wind, before we give our discussion and conclusions in section\,\ref{sec:discussion_conclusions}.

%%%%%%%%%%%%%%%%%%%%%%%%%%%%%%%%%%%%%%%%%%%%%%%%%%%%%%%%%%%%%%

\section{{\sc xwind} model definition}
\label{sec:xwind}

We give here a detailed overview of the calculations that go into out wind line model. Throughout we will use the standard notation for radii, where $r$ is dimensionless gravitational radius and $R$ is in physical units, related via $R = r R_G$ where $R_G = GM/c^2$. This same notation also applies to any other distance measure throughout this paper.

\subsection{Wind Geometry}

\begin{figure*}
    \centering
    \includegraphics[width=0.9\textwidth]{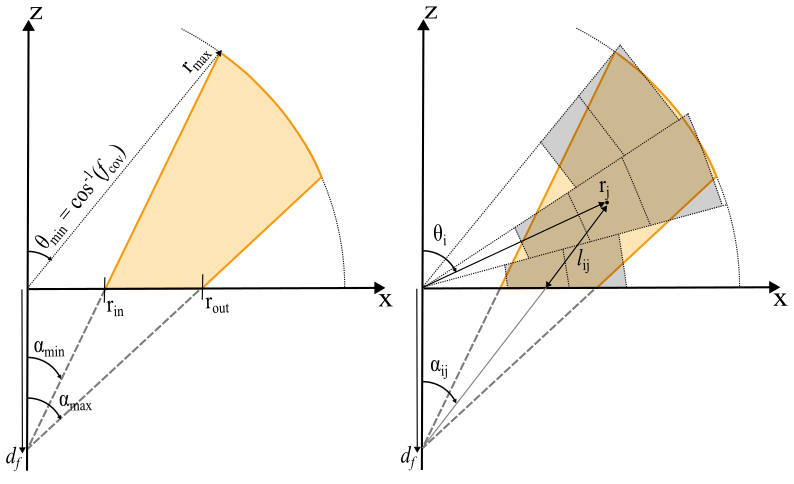}
    \caption{
    \textbf{\textit{Left:}} Definition of our model geometry. The wind is launched between $r_{\rm{in}}$ and $r_{\rm{out}}$, at an angle defined by the distance of the focus below the origin, $d_f$. The covering fraction $f_{\rm{cov}}$ defines the maximum radial extent of the wind $r_{\rm{max}}$ (in spherical polar coordinates), giving a spherical boundary condition. \\
    \textbf{\textit{Right:}} Example of our calculation grid (at significantly lower resolution for clarity). We split the wind into cells of $\cos(\theta)$ with width $d\cos(\theta)$ in the range $\cos(\theta) \in [0, f_{\rm{cov}}]$, and radius $r$ with width $d\log(r)$. The grid in $r$ extends from the inner streamline to either the outer streamline (defined from $r_{\rm{out}}$) or $r_{\rm{max}}$, depending on whichever is smallest. Our wind cells are illustrated as the dark shaded regions. Throughout this paper we will evaluate the wind emission and density at the centre of each cell $\theta_i$ and $r_j$. For completeness, this also defines the length travelled along the streamline $l_{ij}$ at each cell.
    }
    \label{fig:geom_definition}
\end{figure*}

We start by defining the geometry of our wind, \rev{sketched in Fig.\,\ref{fig:geom_definition}}, expanding upon that initially defined in \citet{Sim08}. Throughout we will work in spherical polar coordinates. For simplicity we assume the wind is axisymmetric in azimuth, eliminating any dependence on $\phi$ in \emph{the reference frame of the wind}. 

As in \citet{Sim08} the wind is launched between $r_{\rm{in}}$ and $r_{\rm{out}}$. We consider the wind to consist of a set of streamlines, each with opening angle $\alpha_{ij}$, which all converge at a focal point $d_f$ located on the z-axis directly below the origin. We treat this focal point as an input parameter throughout our calculations. The maximum extent of the wind is then set by the covering fraction $f_{\rm{cov}}$, defined in terms of the total solid angle $f_{\rm{cov}} = \Omega/4\pi$. While this assumes the wind is launched from both sides of the disc, the observer will only see emission originating from the side of the disc facing the observer. This gives a spherical boundary condition of $r_{\rm{max}}$ defined as:

\begin{equation}
    r_{\rm{max}} = \frac{r_{\rm{in}}}{\sin(\theta_{\rm{min}}) - \frac{r_{\rm{in}}}{d_f} \cos(\theta_{\rm{min}})}
\end{equation}

where $\cos(\theta_{\rm{min}}) = f_{\rm{cov}}$ (from the definition of the solid angle).
We stress that this boundary condition is only valid if $\arccos(f_{\rm{cov}}) > \arctan(r_{\rm{in}}/d_f)$, as otherwise the opening angle of the wind is too shallow to ever reach the desired covering fraction. In these instances our model code will throw a warning, and then re-adjust $f_{\rm{cov}}$, before proceeding with the calculation.

We now depart somewhat from the geometry in \citet{Sim08}, here opting to divide our wind into cells in $\theta$ and $r$. We define bins in $\cos(\theta)$, spaced at a constant $d\cos(\theta)=0.002$, in the range $\cos(\theta) \in [0, f_{\rm{cov}}]$. This naturally leads to each wind cell subtending the \emph{same} solid angle $d\Omega = d\phi d\cos(\theta)$ as seen from the central source. When evaluating wind properties we use the centre of each bin, defined as $\theta_i = \cos(\theta_{\rm{edge}}) + 0.5d\cos(\theta)$, where $\theta_{\rm{edge}}$ is the lower edge of the bin.

For each $\theta_i$, we then further subdivide in bins of $r$, geometrically spaced such that $d\log(r)=0.01$ is constant. Here the radial bin edges are defined from the point where the sight-line along $\theta_i$ crosses the streamline launched from $r_{\rm{in}}$ to the point where the sight-line \emph{either} crosses the streamline from $r_{\rm{out}}$ \emph{or} where it crosses $r_{\rm{max}}$; whichever is smallest. As with $\theta$, we evaluate the wind properties at the centre of the bin, defined as $r_j = r_{\rm{edge}} + 0.5\times10^{\log_{10}(r_{\rm{edge}}) + d\log(r)}$.

The wind is now tiled by a set of cells, each with evaluation points $\theta_i$ and $r_j$. This is illustrated in the right side of Fig.\,\ref{fig:geom_definition}. Each of these cells correspond to a streamline launched at $r_{L,ij}$ with opening angle $\alpha_{ij}$, and evaluated at a length $l_{ij}$ along the streamline. These are all given by:

\begin{equation}
    \cos(\alpha_{ij})= \frac{r_j \cos(\theta_i) + d_f}{\big[r_j^2 + 2r_j d_f \cos(\theta_i) + d_f^2\big]^{1/2}}
\end{equation}

\begin{equation}
    r_{L, ij} = d_f \tan(\alpha_{ij})
\end{equation}

\begin{equation}
    \label{eqn:l_stream}
    l_{ij} = \big[r_j^2 + r_{L, ij}^2 - 2r_j r_{L, ij} \sin(\theta_i)\big]^{1/2}
\end{equation}

\subsection{Velocity Profile}
\label{sec:vprof}

The velocity profile can now be easily calculated for each wind cell. 
%We stress that throughout we express the velocity in units of $c$, dropping explicit reference to this in all following equations. 
Starting with the azimuthal velocity $v_\phi$, we assume that at the base of the wind the streamline has the same azimuthal velocity as the disc, taken here to be Keplerian. As material travels out along the streamline, angular momentum is conserved, such that the azimuthal velocity at any point in the wind is \citep{Knigge95}:

\begin{equation}
    \frac{v_{\phi}}{c} = \frac{r_{L}}{r} \sqrt{\frac{1}{r_{L}}}
\end{equation}

Note that here we have dropped explicit reference to the cell indexes $i$ and $j$.

The second component is the outflowing velocity along a streamline. Here we use the velocity profile parametrised in terms of the velocity at infinity $v_{\infty}$ \citep{Knigge95, Sim08, Hagino15, Hagino16, Matzeu22}:

\begin{equation}
    \label{eqn:vl}
    v_l = v_0 + (v_\infty - v_0) \Big(1 - \frac{r_v}{r_v + l} \Big)^\beta
\end{equation}

Here $v_0$ is the initial velocity at the base, which we fix to $0$ throughout. $r_v$ is the velocity scale length, defined as the distance along the streamline where the velocity reaches half of $v_{\infty}$, and $\beta$ is the velocity exponent that determines the rate of acceleration along a streamline. $l$ is the distance along the streamline, previously given in Eqn.\,\ref{eqn:l_stream}.

\subsection{Density Profile}

Given the defined geometry and corresponding velocity profile, we can now estimate the density profile of the wind, following \citet{Matzeu22}. We start by considering a total mass-outflow rate $\dot{M}_w$. It is often easier to express this in terms of the Eddington mass-accretion rate, $\dot{M}_{\rm{Edd}}$, as this naturally scales by the mass of the central object, such that $\dot{m}_w = \dot{M}_w/\dot{M}_{\rm{Edd}}$.

The mass outflow is distributed between $r_{\rm{in}}$ and $r_{\rm{out}}$, with a mass-loss rate per unit area given by $d\dot{M_w}/dA \propto R_L^\kappa$. Here $\kappa$ determines the efficiency of the wind as a function of physical launch radius. For radiatively driven outflows the expectation is that $\kappa \sim -1.3 \to -1$ \citep{Behar09}, which effectively states that the majority of the material is launched from the inner edge of the wind, with the outflow becoming increasingly inefficient as the launch radius increases.

Mass conservation requires that:

\begin{equation}
    \dot{M}_w = 2 \int\limits_S \frac{d\dot{M}_w}{dA} dA 
    = 2 \int\limits_{0}^{2 \pi} d\phi \int\limits_{R_{\rm{in}}}^{R_{\rm{out}}} \frac{d\dot{M}_w}{dA} R dR 
\end{equation}

This is simply the integral across the surface of the disc covered by the wind (i.e the annulus between $r_{\rm{in}}$ and $r_{\rm{out}}$), and the factor of $2$ comes from the disc having two sides. This can be solved analytically, to give:

\begin{align}
    \label{eqn:dMdot_dA}
    \frac{d\dot{M}_w}{dA} &= \frac{\dot{M}_w (\kappa + 2)}{4\pi [R_{\rm{out}}^{\kappa + 2} - R_{\rm{in}}^{\kappa+2}]} R_L^{\kappa} \nonumber \\
    &= \frac{\dot{m}_w \dot{M}_{\rm{Edd}} (\kappa + 2)}{4\pi [r_{\rm{out}}^{\kappa+2} - r_{\rm{in}}^{\kappa+2}]} r_L^{\kappa} R_G^{-2}
\end{align}

\begin{figure*}
    \centering
    \includegraphics[width=\textwidth]{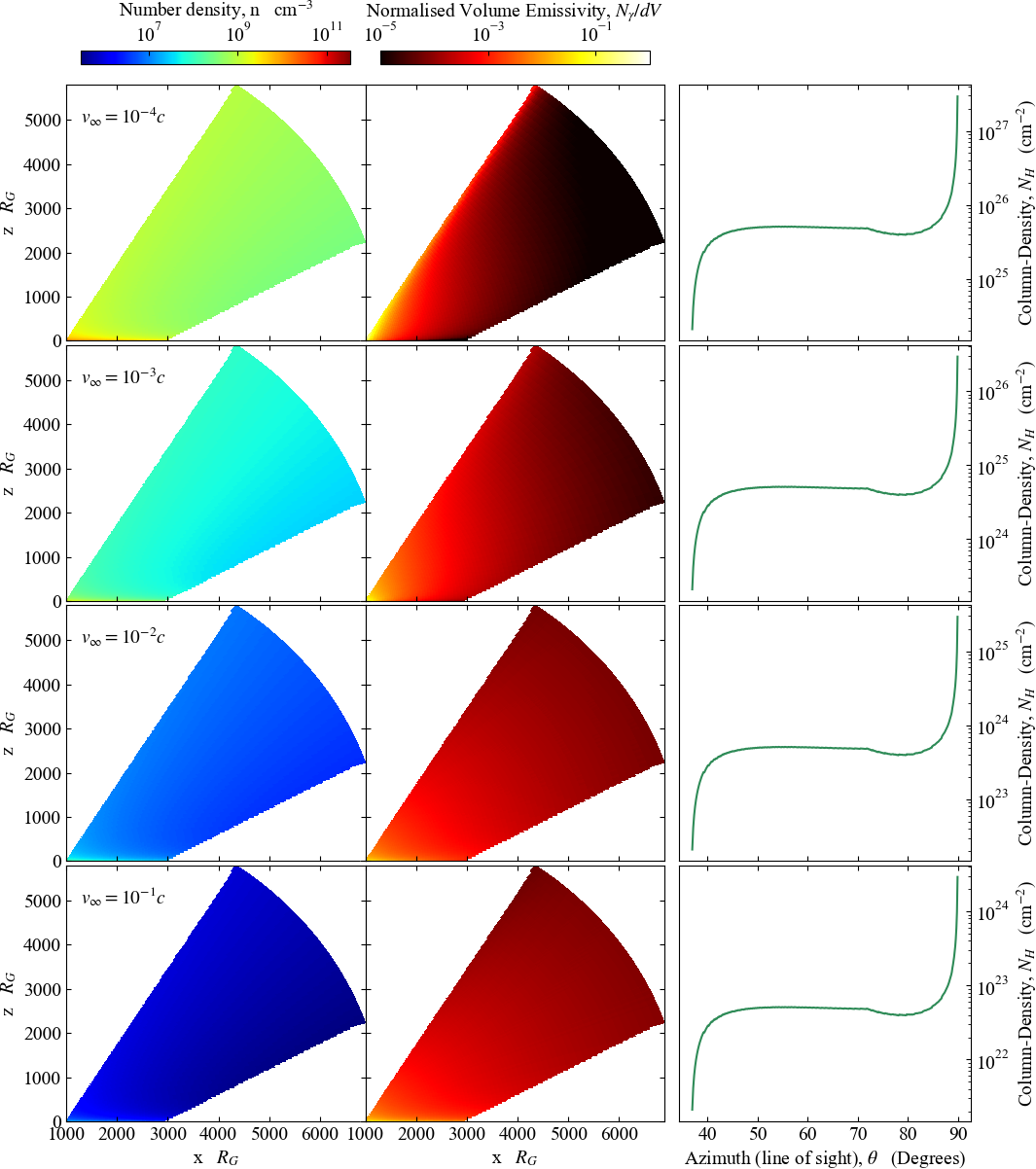}
    \caption{
    Wind density profile (\textbf{\textit{left column}}) and corresponding emissivity profile (\textbf{\textit{middle column}}) for a range of outflow velocities (increasing from top to bottom, value given in top left corner of each row). The \textbf{\textit{right column}} shows the column-density profile for varying lines-of-sight through the winds. These have all been calculated for a wind launched between $r_{\rm{in}} = 1000$ and $r_{\rm{out}} = 3000$, with $d_{f} = r_{\rm{in}}\sqrt{3}$ and $f_{\rm{cov}} = 0.9$. The mass outflow rate has been set to $\dot{m} = 0.1$, for a $10^8\,M_{\odot}$ black hole (where the mass is required to obtain physical units). The velocity profile has been given a scale length of $r_v = 500$ and exponent $\beta = 1$, and the initial density profile (at the base) has $\kappa = -1$, weighting the outflow towards smaller radii. As expected, when the outflow velocity increases, the overall density also reduces. In general, the emissivity is weighted more strongly towards the base of the wind, where the density is highest (hence stronger absorption/fluorescence). For low velocity (high density), the emission will be strongly weighted to the inner edge of the wind, as now the absorption is sufficiently strong that the internal wind regions do not see significant illumination. Our highest velocity examples are chosen for demonstrative purposes rather than an attempt at a realistic BLR wind.
    }
    \label{fig:dens_and_emiss}
\end{figure*}

For a given wind cell, we can write the mass density as $\rho = dM/dV$, where $dV = dA dz$. Further, within a cell the material has a velocity along the streamline $v_l$, such that in time $dt$ it has travelled $dz = v_l dt$. Hence we can write $dV = v_l dt dA$, such that $\rho = d\dot{M}_w/(v_ldA)$. The number density in a wind cell is then given by:

\begin{equation}
    n = \frac{\rho}{m_I} = \frac{1}{m_I v_l} \frac{d\dot{M}_w}{dA}
\end{equation}

where $m_I \simeq 1.23 m_p$ is the ion mas, which we choose to write in terms of the proton mass $m_p$. the left column of Fig.\,\ref{fig:dens_and_emiss} demonstrates the density profile for a wind launched at BLR like distances over a range of outflow velocities.

\subsection{Calculating the emissivity}
\label{sec:calc_emiss}

An important aspect in the shape of the emission line is the relative emissivity across the wind \rev{volume}. For a photo-ionised line this can be calculated from an incident spectrum and wind density profile. We give here an overview of our methodology for this, based on the neutral Fe-K$\alpha$ line. We stress, that the below calculations give self-consistent solutions to the Fe-K$\alpha$ line strength, however for other emission lines this is not the case. Hence within {\sc xwind}, we provide two submodels: {\sc xwindline}, where the normalisation should remain free for generic use, and {\sc xwindfe}, where the normalisation is self-consistently calculated for \emph{neutral} Fe-K$\alpha$ and the rest-frame emission assumes the 7-Lorentzian approximation from \citet{Holzer97}. Internally, the calculations are identical, as {\sc xwindline} uses the emissivity calculated assuming neutral Fe-K$\alpha$. We also use {\sc xwindline} to extract a convolution kernel, {\sc xwindconv}, allowing for the application of wind broadening to any spectral-shape, including continuum. \rev{Throughout we neglect the Compton shoulder in {\sc xwind}, as the typical wind densities are too low for this to be important; with the exception perhaps at the base of the wind.}

We base our calculations on the analytic approach given in \citet{Yaqoob01}, updated in \citet{Tzanavaris23}. Firstly, the wind will see an incident spectrum, $N(E)$, originating from the central regions of the accretion flow. The relevant part of the spectrum depends somewhat on the line species being calculated, due to strongly differing ionisation potentials. However, this is too complex for our approach. Instead we assume that the \emph{relative} emissivity across the wind volume is more or less identical for each line species (i.e that for all lines it traces density), and only the absolute (i.e normalisation) emissivity differs. Thus, internally we calculate the absolute emissivity for neutral Fe-K$\alpha$, and simply apply the same emissivity for any other line.

The neutral Fe-K$\alpha$ line is induced by photons above $E_K \simeq 7.1$\,keV. In this energy range the intrinsic emission is dominated by inverse Compton scattering, giving a power-law like spectrum \citep{Haardt93, Haardt94}. Hence, we fix the incident spectrum to a power-law of the form $N(E) = N_0 E^{-\Gamma}$, photons s$^{-1}$ cm$^{-2}$ keV$^{-1}$. In {\sc xwindfe} $N_0$ is the number of photons s$^{-1}$ cm$^{-2}$ at $1$\,keV. Following \citet{Tzanavaris23}, the emitted Fe-K$\alpha$ photon flux from a constant column-density material with a solid angle $\Omega$ is:

\begin{equation}
    \label{eqn:Ne_fek}
    N_{\rm{Fe-K}\alpha} = \frac{\Omega}{4\pi} y \int\limits_{E_K}^{\infty} N(E) \Big\{ 1 - \exp(-\sigma_{\rm{Fe-K}}(E) A_{\rm{Fe}} N_H \Big\} dE  
\end{equation}

where $y \sim 0.3$ is the fluorescent yield, $A_{\rm{Fe}} \simeq 4.86\times 10^{-5}$ \citep{Anders89} is the iron abundance, $N_H$ is the column-density of the absorbing material, and $\sigma_{\rm{Fe-K}}$ is the iron K-shell absorption cross-section. The cross-section decays exponentially with energy above the absorption edge, such that $\sigma_{\rm{Fe-K}}(E) = \sigma_0 (E/E_K)^{-\alpha}$ where $\alpha \simeq 2.67$ and $\sigma_0 \simeq 3.37 \times 10^{-20}$\,cm$^{2}$ \citep{Murphy09, Tzanavaris23}.

Of course, our wind does not have a constant column-density. Instead this needs to be calculated separately for each wind cell. Making the simplifying assumption that the density within a single wind cell is roughly constant, we can write for the column-density in cell $ij$: $N_{H, ij} \simeq n_{ij} \Delta R_j$, where $\Delta R_j = r_j R_G d\log r$ is the radial width of the wind cell, and is directly calculated from the model geometry.

As we integrate radially along a sight-line $\theta_i$, not only does the column-density evolve, but also the spectrum incident into the wind cell. For a given cell at ($\theta_i$, $r_j$), the incident spectrum $N(E)$ will be slightly absorbed by the cell at ($\theta_i$, $r_{j-1}$). Hence, the fluorescence from each cell is not simply calculated directly from Eqn.\,\ref{eqn:Ne_fek}. Instead, we need to take into account the effect from \emph{all} radially preciding cells on the incident spectrum, such that for a wind cell $ij$ at ($\theta_i$, $r_j$), the incident spectrum is given by:

\begin{equation}
    \label{eqn:Ne_cell}
    N_{ij}(E)= N(E) \prod\limits_{k=1}^{j-1} \exp(-\sigma_{\rm{Fe-K}}(E) A_{\rm{Fe}} n_{ik} r_k d\log r R_G)
\end{equation}

and so the fluorescence from the wind cell at $ij$ is:

\begin{equation}
    \label{eqn:Ng_em}
    \begin{split}
    N_{\gamma, ij} = \frac{d\cos(\theta)d\phi}{4\pi} y \int\limits_{E_K}^{\infty} \Big(N_{ij}(E) - N_{i,j+1}(E)\Big) dE \\
    \rm{photons\,\,s^{-1}\,\,cm^{-2}}
    \end{split}
\end{equation}

which now also takes into account the solid angle of each wind cell as seen from the central source. In Eqn.\,\ref{eqn:Ne_cell}, all factors of $M$ cancel (by expanding $R_G$ and $\dot{M}_{\rm{Edd}}$), such that the total fluorescence and line profile is fully independent of mass.

\subsection{Calculating the overall line profile}
\label{sec:line_prof}

The previous subsection only gives the emission in the co-moving frame of the wind. In the observes frame, the emission will be smeared out into a characteristic line shape. Material moving away/towards the observer will broaden the line due to Doppler shifts, the bulk outflowing motion of the wind will give a net blueshift, and relativistic length contraction and time-dilation will lead to a boosting effect in the emission originating from material travelling towards the observer. These are all very well known effects, which have previously been calculated for accretion discs \citep{Chen89, Fabian89}. In this section we give a generalisation to a 3D velocity profile, though \emph{only in the weak field limit}. 
%For completeness, we give here a full derivation. 

%We start by calculating the energy shift between the observed and emitted frames:

%\begin{equation}
%    \frac{E_{\rm{em}}}{E_{\rm{obs}}} = 1+z
%\end{equation}
 
%where $z$ is the net redshift. From \citet{Luminet79} we can write the emitted photon energy as vector 4-product between the 4-velocity of the emitting material, $u^{\alpha}$, and the 4-momentum of the photon travelling to the observer, $p^{\alpha}$, such that $E_{\rm{em}} = u_\alpha p^\alpha$.

In general, the energy of the emitted photon is given by the projection of the photon 4-momentum, $p^{\alpha}$, onto the 4-velocity of the emitting material, u$^{\alpha}$, such that $E_{\rm{em}} = u_{\alpha}p^{\alpha}$ \citep{Cunningham75, Luminet79}. This will give the usual relation for the special relativistic Doppler shift in a flat space-time; as considered here.

The 4-velocity of the emitting material is simply given as $u_\alpha = g_{\alpha \beta} u^{\beta} = \gamma(c, -\mathbf{v_w})$ where $\mathbf{v_w}$ is the 3-velocity profile of the wind, $\gamma = (1 - (v_w/c)^2)^{-1/2}$ is the Lorentz factor of the wind material, and $g_{\alpha \beta}$ is the Minkowski metric. The wind 3-velocity components were previously calculated in section\,\ref{sec:vprof}, such that in Cartesian coordinates:

\begin{equation}
    \begin{split}
    u_\alpha = \gamma \Big\{ c, &-v_l\sin(\alpha)\cos(\phi) + v_\phi \sin(\phi),  \\
    &-v_l\sin(\alpha)\sin(\phi) - v_\phi \cos(\phi), -v_l\cos(\alpha) \Big\}
    \end{split}
\end{equation}

The 4-momentum of a photon is simply given by $p^\alpha = E_\gamma/c(1, \mathbf{\hat{n}}_\gamma)$, where $\mathbf{\hat{n}}_\gamma$ is a unit vector describing the direction of the photon. Energy is conserved along a trajectory, and so here $E_\gamma = E_{\rm{obs}}$ \citep{Cunningham75}. Further, since we assume the photon travels in straight lines, the photon direction is simply $\mathbf{\hat{n}}_\gamma = (\sin(i), 0, \cos(i))$ for an observer located in the x-z plane at an inclination $i$ from the z-axis. Hence, we arrive at the ratio between observed and emitted frame photon energies, \emph{considering special relativity only}:

\begin{equation}
    \begin{split}
    \frac{E_{\rm{em}}}{E_{\rm{obs}}} = \frac{u_\alpha p^\alpha}{E_{\rm{obs}}} = &\gamma \Bigg\{1 -
    \frac{1}{c} \bigg[v_l\cos(\alpha)\cos(i) + \\
    &\sin(i) \Big(v_l\sin(\alpha)\cos(\phi) - v_\phi \sin(\phi) \Big) \bigg] \Bigg\}
    \end{split}
\end{equation}

As expected, this is in the form of the special relativistic Doppler shift, except here it is decomposed into individual velocity components. Also factoring in the gravitational redshift, calculated in the Schwarzchild metric, gives:

\begin{equation}
    \label{eqn:Efrac}
    \begin{split}
    \frac{E_{\rm{em}}}{E_{\rm{obs}}} =  &\gamma \bigg(1 - 
    \frac{2}{r} \bigg)^{-\frac{1}{2}}\Bigg\{1 -
    \frac{1}{c} \bigg[v_l\cos(\alpha)\cos(i) + \\
    &\sin(i) \Big(v_l\sin(\alpha)\cos(\phi) - v_\phi \sin(\phi) \Big) \bigg] \Bigg\}
    \end{split}
\end{equation}

The above only gives the relation between emitted and observed photon energies. To relate this to an observed photon flux, we take advantage of the fact that $N_\gamma/E^3$ is Lorentz invariant \citep{Lindquist66, Cunningham75}, such that:

\begin{equation}
    \label{eqn:Nobs}
    \begin{split}
    N_{\gamma, \rm{obs}} = \left( \frac{E_{\rm{em}}}{E_{\rm{obs}}} \right)^{-3} N_{\gamma, \rm{em}} \\
    \rm{photons}\,\,s^{-1}\,\,\rm{cm}^{-2}
    \end{split}
\end{equation}

where $N_{\gamma, \rm{em}}$ is calculated from Eqn.\,\ref{eqn:Ng_em}.

The total line-profile comes from integrating over the wind volume. However, as we do not know a priori what wind cells correspond to what specific energy shift, it is simpler to iterate over each cell, calculate the photon flux and energy shift from Eqns.\,\ref{eqn:Ng_em}, \ref{eqn:Efrac}, and \ref{eqn:Nobs}, adding the output photon flux to the relevant energy bin.

Throughout we use an internal fractional energy grid in $E/E_0$, with bins evenly spaced from $E/E_0 = 0.2$ to $4.4$ with $\Delta(E/E_0) = 2\times10^{-4}$. This allows for maximum velocity widths up to $0.9c$, which is more than adequate for any realistic wind. It also gives an internal energy resolution of $1.2$\,eV at $6$\,keV, oversampling that of \xrism-resolve ($\sim 5$\,eV at $6$\,keV, \citealt{Kelley25_resolve}). 

As a sanity check, we apply our above methodology to a disc geometry, allowing for a direct comparison to {\sc diskline} \citep{Fabian89}. This shows generally very good agreement for large radii, but begins to deviate systematically below $r \sim 50$ (see Appendix\,\ref{app:diskline_comp}, Fig.\,\ref{fig:disklineComp}). This is expected, since {\sc diskline} considers a treatment using the Schwarzchild metric (though no light-bending). \textit{As such, we do not recommend the use of our model code for small launch radii.}

\section{Model Properties}
\label{sec:mod_properties}

\subsection{Emission Line Profiles}

\begin{figure*}
    \centering
    \includegraphics[width=0.99\textwidth]{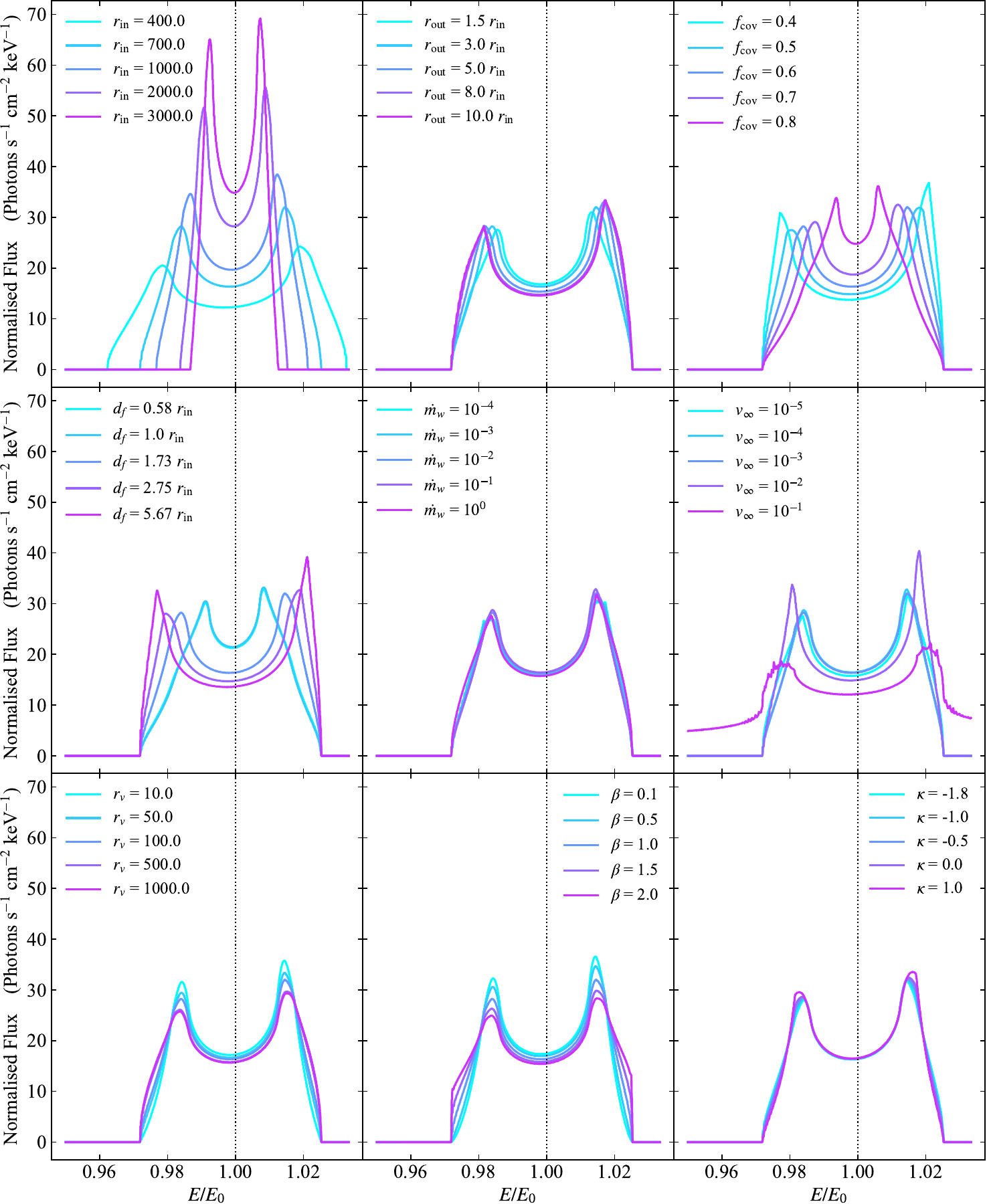}
    \caption{
    Example emission line profiles and their evolution with varying wind parameters. These have all been re-normalised to have the same total flux, and have all been calculated for an observed inclination of $45$\,deg. Each panel shows the effect of varying an individual parameter, given in the top left corner of each panel. When not being varied, parameters are fixed at: $\dot{m}_w = 10^{-2}$, $r_{\rm{in}} = 700$, $r_{\rm{out}} = 3r_{\rm{in}}$, $d_f = \sqrt{3} r_{\rm{in}}$, $f_{\rm{cov}} = 0.6$, $v_{\infty} = 10^{-3}$, $r_v = 100$, $\beta = 1$, $\kappa=-1$. These are all viewed at an observers inclination of $i=45$\,deg. The parameter choices for $d_f$ are chosen such that they correspond to wind opening angles on the inner edge of $\alpha_{\rm{min}} = 60^{o}, 45^{o}, 30^{o}, 20^{o},$ and $10^{o}$ (see Fig.\,\ref{fig:geom_definition} for a definition of the opening angle). 
    }
    \label{fig:lin_profiels}
\end{figure*}

We illustrate the model emission line profiles in Fig.\,\ref{fig:lin_profiels}. We illustrate the effects by varying each parameter individually, while keeping the others fixed. The underlying fixed model is calculated for $\dot{m}_w = 10^{-2}$, $r_{\rm{in}} = 1000$, $r_{\rm{out}} = 3 r_{\rm{in}}$, $d_f=\sqrt{3} r_{\rm{in}}$, $f_{\rm{cov}} = 0.6$, $v_{\infty} = 10^{-3}$, $r_v = 100$, $\beta=1$, and $\kappa = -1$. All line profiles have been normalised to unity, in order to highlight changes in the shape, and are observed at an inclination of $45$\,degrees.

To start, there is a clear trend in that the vast majority of the line profiles appear rather sharp. This is not surprising when we examine the emissivity profiles in Fig.\,\ref{fig:dens_and_emiss}. In general the emission is dominated by the inner edge at the base of the wind, as this is where both the density and incident flux is the strongest. At larger distances along the streamline, the density drops off, such that the absorption becomes less efficient, resulting in fewer photons available to re-emit. Alternatively, for lines of sight passing through a dense region (e.g at the base), the optical depth will be high, and so the incident continuum is mostly absorbed before it reaches the outer edge of the wind. This is clear in Fig.\,\ref{fig:lin_profiels}, which shows almost no change in the line profile when changing $r_{\rm{out}}$.

%The main parameters of interest, which have a noticeable impact on the line profile, are then $r_{\rm{in}}$, $f_{\rm{cov}}$, $d_f$, and $v_{\infty}$. There are also some second order effect from changing the velocity profile parameters, $r_v$ and $\beta$, however it is unlikely that this would be detectable.

In terms of understanding the physical origin of the wind, the key parameters are $r_{\rm{in}}$, $v_{\rm{\infty}}$ and $\dot{m}_w$. The combination of $r_{\rm{in}}$ and $v_{\infty}$ should inform the driving mechanism of the wind. Especially radiation driven outflow models, e.g UV line-driven winds \citep{Proga00, Proga04}, FRADO models \citep{Czerny11, Naddaf21}, tend to have a specific radial range over which they can exist, as they strongly depend on the local disc SED, as well as containing predictions on the outflow velocity. As these both have significant impact on the line profile, they should be possible to constrain.

\begin{figure*}
    \centering
    \includegraphics[width=\textwidth]{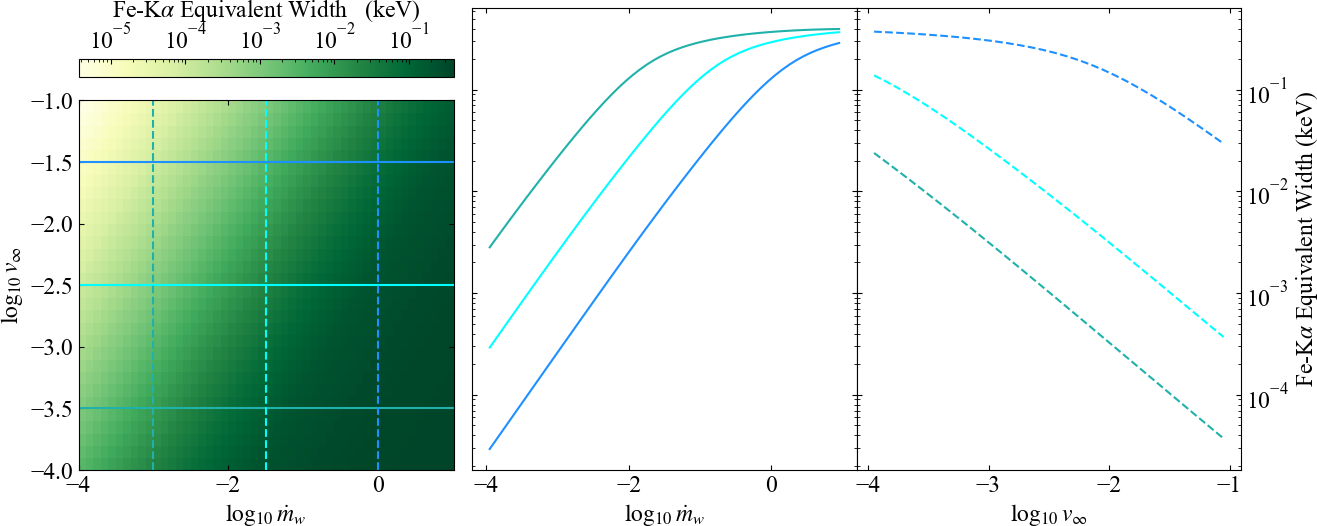}
    \caption{
    \textbf{\textit{Left:}} Colour-plot showing the equivalent width of the Fe-K$\alpha$ emission line for varying $\dot{m}_w$ and $v_{\infty}$. The darker the shade of green, the larger the equivalent width. The vertical/horizontal solid/dashed lines correspond to the slices used to extract the curves of growth on the right as a function of $\dot{m}_w$/$v_{\infty}$ \\
    \textbf{\textit{Right:}} Curves of growth as a function of $\dot{m}_w$ (left panel) and $v_{\infty}$ (right panel). As $\dot{m}_w$ increases, so does the equivalent with, until eventually the line saturates due to the wind going optically thick. Conversely, increasing $v_{\infty}$ will reduce the equivalent width, as this reduces the overall density of the wind (see Fig.\,\ref{fig:dens_and_emiss}) leading to a lower degree of absorption and thus fewer photons available to re-emit as Fe-K$\alpha$.
    }
    \label{fig:feka_ew}
\end{figure*}

The behaviour of the line profile with $v_{\infty}$ also reveals a key effect of calculating the density profile via mass-conservation. As one reduces the velocity, the overall density of the wind increases. In addition it appears from Fig.\,\ref{fig:dens_and_emiss} that the moderate to high density regions also move further up the streamline. This will then increase the fraction of the emission originating from larger distances along a given streamline within the wind, which in turn traces a smaller angular velocity due to the conservation of angular momentum. The net effect is that the line profile begins to `fill in' towards the centre. 

\subsection{Fe-K$\alpha$ Equivalent Width}
\label{sec:xwind_EW}

A key advantage of our approach is that it gives a self-consistent normalisation to the Fe-K$\alpha$ emission line for a given wind density profile and incident X-ray spectrum (as also outlined in \citealt{Tzanavaris23}). This depends strongly on both the mass-outflow rate, $\dot{m}_w$ and outflow velocity profile (governed by $v_{\infty}$), as these determine the density profile. We demonstrate this in Fig.\,\ref{fig:feka_ew}. This shows on the left the expected equivalent width of the Fe-K$\alpha$ line for a grid of $\dot{m}_w$ and $v_{\infty}$. On the right we show curves of growth with $\dot{m}_w$ and $v_{\infty}$ respectively. These are calculated for the same base model as the line-profiles in Fig.\,\ref{fig:lin_profiels}. The equivalent widths are calculated with respect to the incident continuum (a power-law, see section\,\ref{sec:calc_emiss}) at (rest-frame) 6.4\,keV.

These behave as expected. Increasing $\dot{m}_w$ leads to more material in the wind, and so naturally the density increases. Higher density leads to increased absorption of the incident spectrum, which in turn gives a higher number of available photons to re-emit. The perhaps more interesting effect is the saturation of the equivalent width once $\dot{m}_{w}$ reaches a certain value. This is an effect of the wind going optically thick, and so now the limiting factor to the total fluorescence depends on the incident continuum rather than the wind density profile. However, increasing the normalisation of the incident continuum does not lead to a higher equivalent width, as this is always measure relative to the continuum. Hence, we predict a rather stringent upper limit on the expected Fe-K$\alpha$ equivalent width from a wind of $EW \sim 5\times 10^{-1}$\,keV (for a covering fraction of 0.6. The exact value will naturally also depend somewhat on $f_{\rm{cov}}$). This also becomes clear from Fig.\,\ref{fig:dens_and_emiss} which shows how the emission becomes strongly concentrated to the inner edge of the wind when the density is sufficiently high, such that the incident spectrum is highly absorbed before reaching the inner/outer regions (top panel).

\section{Application to XRISM data - a wind on the inner edge of the BLR in NGC\,4151?}
\label{sec:application_xrism}

We demonstrate our model on a \xrism\ observation of NGC\,4151, taken during the performance verification (PV) phase. NGC\,4151 was observed five times during the PV phase \citep{XRISM24_NGC4151, Xiang25}. We pick the longest observation to maximise signal-to-noise (obsID\,300047020, observation date 18.05.2025, exposure time 103.7\,ks, presented previously in \citealt{Xiang25}).

Before constraining the line-profile, we also fit the broad-band continuum, using simultaneous \xrism-xtend \citep{Noda25_xtend} and NuSTAR data. Below we give details on the data reduction, broad-band spectral fit, and finally our fit to the \xrism-resolve \citep{Kelley25_resolve} spectrum of the Fe-K$\alpha$ complex.

For all subsequent analysis, we use the reduction pipelines included in {\sc heasoft}\,v.6.63, and {\sc xspec}\,v.12.15.1 \citep{Arnaud96}. All errors are quoted at 90\,\% confidence.

\subsection{\xrism-data reduction}

We reduce both the \xrism-resolve and \xrism-xtend data, using the standard \xrism\ reduction pipeline along with the corresponding most recent calibration files from the CALDB database (as of writing corresponding to the resolve and xtend calibration files from 15. September 2025, and the general files from 15 November 2024). We give here a brief overview of the reduction process, as also outlined in \citet{XRISM24_NGC4151}.

We start by running {\sc xapipeline}, which generates the cleaned event files for both resolve and xtend. Next, focusing in \xrism-resolve, we use {\sc xselect} to extract the source spectrum, filtering such that we \emph{only} extract high-resolution primary (Hp) events. We also filter out periods of high particle background, and data from pixel\,27 (as recommended due to a significant variation in gain). Using {\sc rslmkrmf} we then generate a RMF file, setting the size option to large (L) in order to accurately give the response around our region of interest (5-7\,keV), but without the significant increase in file size that comes with including the electron loss continuum (the x-large option). For the ancillary response (ARF) file we first generate an exposure map using {\sc xaexpmap}, before then generating the ARF using {\sc xaarfgen}. Here we set the source type to a point-source, and run the ray-tracing of reflected and transmitted photons through the detector using 300000\,photons.

For \xrism-xtend we extract a source spectrum from a rectangular region with dimensions $2.3'' \times 5.0''$ centred on the source. We then extract a background spectrum using the same extraction region, but on a portion of sky with no source contribution. The RMF file is generated using {\sc xtdrmf}, and an ARF file in an identical manner to \xrism-resolve (but with the instrument set to xtend).

For both spectra, we bin the data such that each bin contains \emph{at least} 25\,counts, allowing for the use of $\chi^2$ statistics during the fitting procedure.

\subsection{NuSTAR data reduction}

Simultaneous with the \xrism\ observation, NGC\,4151 was observed by NuSTAR (obsID\,60902010004, presented previously in \citealt{Xiang25}). We combine these data with the \xrism-xtend data, to extend our broad-band continuum fit to 90.0\,keV.

We use the standard {\sc nupipeline} tool to generate the cleaned event files, with CALBD calibration files corresponding to the most recent update as of writing (released 06 October 2025). After generating the cleaned event files, we run {\sc nuproducts} to give both the source and background spectral files, the response file, and the ancillary response file; for both FPMA and FPMB modules. The source spectrum is extracted using a circular annulus with a radius of $120''$ centred on the source, while the background is extracted from an identical annulus in an empty patch of sky. As with the \xrism\ spectra, we re-bin the source spectrum to contain \emph{at least} 25\,counts per bin.

\subsection{Broad-band spectral fit}

Before constraining the profile of the Fe-K$\alpha$ complex we require an estimate of the underlying continuum. This is for two reasons. The first is that the intrinsic source continuum flux above $\sim 7.1$\,keV determines the absolute strength of the Fe-K$\alpha$ fluorescence line (see Eqn.\,\ref{eqn:Ne_fek}). The second is that broad features in the line profile are highly sensitive to the shape of the underlying continuum; especially if these features are weak.

\begin{figure}
    \centering
    \includegraphics[width=\columnwidth]{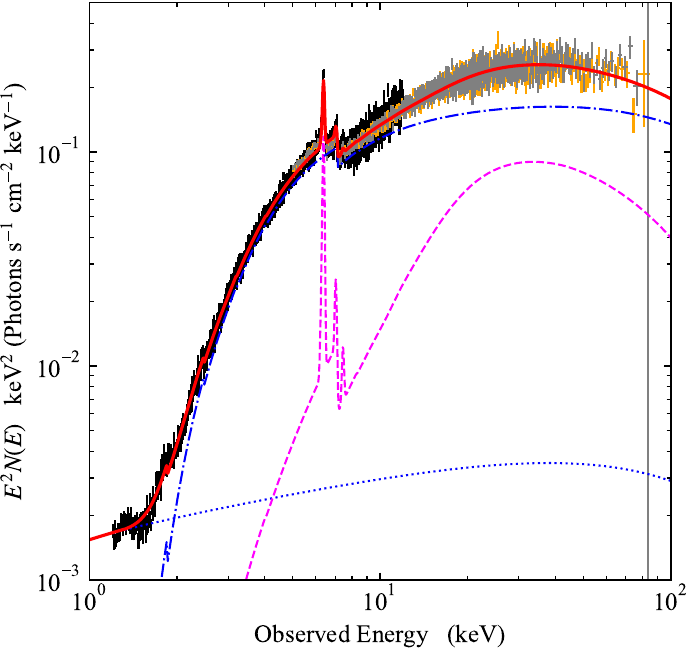}
    \caption{Best fit broad-band continuum, using combined \xrism-xtend (black crosses) and NuSTAR (orange/grey crosses for FPMA/B respectively). The solid red line shows the total spectral model, with the dotted/dashed lines showing the individual components. These are: the scattered continuum (dotted blue), reflection spectrum (dashed magenta), and the primary continuum (dashed-dotted blue). The plotted NuSTAR data have been corrected by a cross-calibration constant used to account for instrumental differences between \xrism-xtend and NuSTAR. The data have been re-binned slightly for clarity.}
    \label{fig:continuumFit}
\end{figure}

Following Noda et al. (in prep) we construct a phenomenological model for the underlying continuum, using here the combined \xrism-xtend and NuSTAR data. At low energies below $\sim 1.5$\,keV the \xrism-xtend spectrum is dominated by a complex scattered continuum. Hence, we limit the \xrism-xtend energy range to $1.2-12$\,keV during the fitting, where we note that part of the scattered continuum is included in order to estimate its contribution to the overall flux level. For NuSTAR, we consider the energy range between 5.0-90.0\,keV; giving a total combined band-pass of 1.2-90\,keV. The total source spectrum is shown in Fig.\,\ref{fig:continuumFit}.

To describe the primary continuum we use a redshifted cut-off power-law ({\sc zcutoffpl} in {\sc xspec}), subject to absorption from the host galaxy ({\sc tbabs} \citealt{Wilms00}) and also partial covering absorption intrinsic to the AGN itself (e.g from a wind) ({\sc pcfabs}). We then include a component originating from cold reflection off the disc ({\sc pexmon} \citealt{Magdziarz95, Nandra07}) convolved with Gaussian broadening ({\sc gsmooth}) to account for the iron emission and Compton hump. Here we fix the observer inclination at 60\,degrees and all abundances to solar abundances. Finally, we include an additional cut-off power-law, with photon index and cut-off energy tied to the primary continuum, to describe the scattered continuum. The total model is then also absorbed by our galaxy, where we fix $N_{H}=2\times 10^{20}$\,cm$^{-2}$.
The final {\sc xspec} model is then: {\sc tbabs$_{\rm{MW}}$ * (zcutoffpl$_{\rm{scatt}}$+ tbabs$_{\rm{int}}$*(pcfabs*zcutoffpl$_{\rm{primary}}$ + gsmooth*pexmon))}. We also include a cross-calibration constant, applied to the NuSTAR data, to account for small calibration differences between \xrism-xtend and NuSTAR.

This gives a moderate fit to the broad-band spectrum, with $\chi^2_{\nu} = 4876/3760 = 1.3$. The increase in $\chi^2$ is predominantly being driven by complexity in the scattered continuum, and in the region $5-8$\,keV where details in the Fe-K$\alpha$ profile not captured by this model as well as absorption lines (as seen in \citealt{Xiang25}) not included here drive significant residuals. As these are a known effect, and since the broad-band continuum shape is otherwise well constrained, we continue with this as our underlying base continuum for our more detailed analysis of the \xrism-resolve spectrum in the next subsection. The best fit continuum model is shown in Fig.\,\ref{fig:continuumFit}, with corresponding fit parameters shown in Table\,\ref{tab:continuumFit}. The errors are estimated at 90\,\% confidence from a MCMC chain using 8 walkers and 160000 steps.

{\renewcommand{\arraystretch}{1.5} %Adjust table spacing
\begin{table}
    \centering
    \begin{tabular}{c c c} 
   \textbf{Component} & \textbf{Parameter (Unit)} & \textbf{Value} \\ 
   \hline 
   \hline 
   {\sc tbabs$_{\rm{MW}}$} & $N_{H}$ ($10^{22}$\,cm$^{-2}$) & $0.02$ \\ 
   \hline 
   {\sc zcutoffpl$_{\rm{scatt}}$} & $\Gamma$ & $1.71^{+0.02}_{-0.01}$ \\ 
      & $E_{\rm{cut}}$ (keV) & $132^{+18}_{-12}$ \\ 
      & Norm. & $0.00165^{+4e-05}_{-2e-05}$ \\ 
   \hline 
   {\sc tbabs$_{\rm{Int}}$} & $N_{H}$ ($10^{22}$\,cm$^{-2}$) & $6.2^{+0.2}_{-0.1}$ \\ 
   \hline 
   {\sc pcfabs} & $N_{H}$ ($10^{22}$\,cm$^{-2}$) & $14.7^{+1.4}_{-0.5}$ \\ 
   & $f_{\rm{cov}}$ & $0.59^{+0.01}_{-0.02}$ \\ 
   \hline 
   {\sc zcutoffpl$_{\rm{primary}}$} & $\Gamma$ & $=\Gamma_{\rm{scatt}}$ \\ 
      & $E_{\rm{cut}}$ (keV) & $=E_{\rm{cut}, {scatt}}$ \\ 
      & Norm. & $0.077^{+0.004}_{-0.002}$ \\ 
   \hline 
   {\sc gsmooth} & $\sigma_{6\,keV}$ & $0.044^{+0.005}_{-0.008}$ \\ 
    & Index & 1 \\ 
   \hline 
   {\sc pexmon} & $\Gamma$ & $=\Gamma_{\rm{scatt}}$ \\ 
   & $E_{\rm{cut}}$ (keV) & $=E_{\rm{cut}, \rm{scatt}}$ \\ 
   & Relative Reflection & -1 \\ 
   & Abund. & 1 \\ 
   & $A_{\rm{Fe}}$ & 1 \\ 
   & Inc. (deg) & 60 \\ 
   \hline 
   Const$_{\rm{NuSTAR}}$ & & $1.072^{+0.005}_{-0.005}$ \\ 
   \hline 
   & Redshift & 0.003326 \\ 
   \hline 
   \hline 
   $\chi^2$/d.o.f & & 4876.56/3760 = 1.3 \\ 
   \hline 
\end{tabular} 

    \caption{
    Best fit parameters to the broad-band fit of the \xrism-xtend and NuSTAR data, using the model: {\sc tbabs$_{\rm{MW}}$ * (zcutoffpl$_{\rm{scatt}}$+ tbabs$_{\rm{int}}$*(pcfabs*zcutoffpl$_{\rm{primary}}$ + gsmooth*pexmon))}. These have all been calculated from an MCMC chain, in {\sc xspec}, using 8 walkers and 160000 steps. The errors correspond to 90\,\% confidence. Parameter values prefixed with an $=$ are tied to the given component. The constant is included to account for calibration differences between \xrism-xtend and NuSTAR.
    }
    \label{tab:continuumFit}
\end{table}
}

\subsection{Model application to \xrism-resolve}

\begin{figure*}
    \centering
    \includegraphics[width=\textwidth]{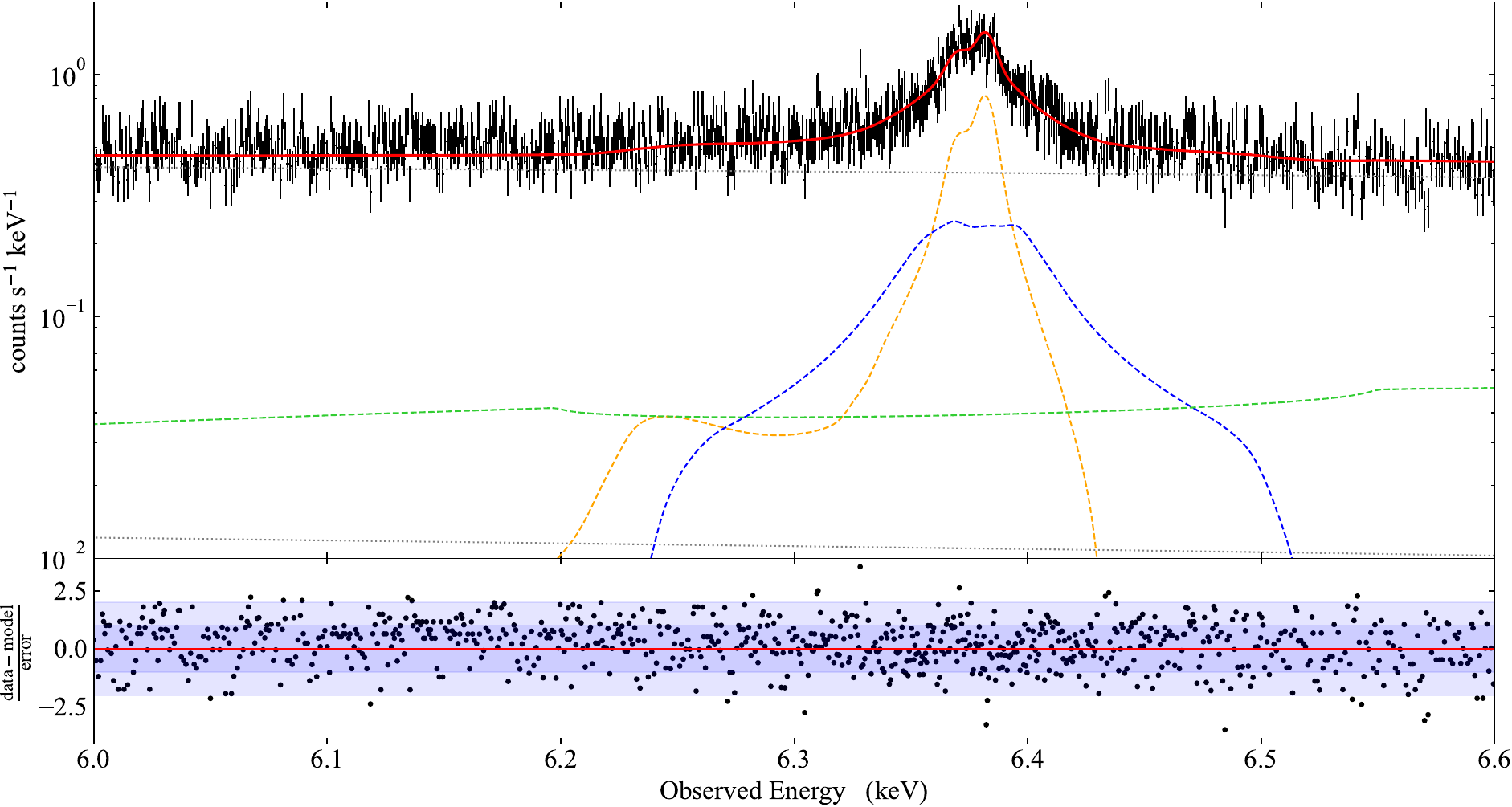}
    \caption{
        \textbf{\textit{Top:}} Best fit model to \xrism-resolve data (black points). The solid red line shows the total model, while the dashed and dotted lines show the components. These are: scattered and primary continuum (grey dotted lines), {\sc mytorus} emission line from Compton thick slow moving material (orange dashed line), {\sc xwindfe} emission line for a wind inwards of the BLR (blue dashed line), and {\sc diskline} emission line from the inner disc (greend dashed line).
        \textbf{\textit{Bottom:}} Data-model residuals, normalised by the data errorbar. The shaded blue bands show the 1$\sigma$ (dark) and 2$\sigma$ (light) deviation.
    }
    \label{fig:resolve_fit}
\end{figure*}

We now focus our analysis onto the \xrism-resolve spectrum. Specifically, we focus on the shape of the Fe-K$\alpha$ complex, and so limit the spectral fit to the energy range $5.0-6.6$\,keV (in the observed frame).

For the underlying continuum we take the model from the broad-band spectral fit, fixing all parameters to those in Table\,\ref{tab:continuumFit}. We make one change, removing the {\sc rdblur*pexmon} component, as we are now focusing on a detailed fit of the Fe-K$\alpha$ complex. \citet{XRISM24_NGC4151} showed three distinct components to the Fe-K$\alpha$ line, which we adapt here. We start with a {\sc mytorus} \citep{Yaqoob12} component, convolved with {\sc rdblur} \citep{Fabian89}, to account for the narrow Torus emission. Specifically, we use the updated {\sc mytorus} line emission files (mytl\_V000HLZnEp000\_v01.fits) from \citet{Yaqoob24} as these include the intrinsic Fe-K$\alpha$ profile from \citet{Holzer97}. In {\sc rdblur} we fix the emissivity index to -2, in order to account for the non-disc geometry of the torus, and set a lower limit of $10^4\,R_G$ for the inner radius. We fix the outer radius at $10^7\,R_G$.

Next we add a {\sc xwindfe} component, to account for the moderately broad BLR scale emission; the main focus of this analysis. We stress that {\sc xwindfe} will self-consistently calculate the equivalent width of the line for a given input spectrum, and also includes the intrinsic \citet{Holzer97} profile for the rest-frame emission (see section\,\ref{sec:xwind} for a full model description). Hence we tie the input spectrum for {\sc xwindfe} to that of the primary continuum in the broad-band fit. We also fix a number of the geometric parameters in order to simplify the fit. Firstly, we set the outer launch radius to $r_{\rm{out}} = 1.5r_{\rm{in}}$, since we are aiming to describe a wind on the inner edge of the BLR, and so this will in effect limit the radial extent of the wind. Secondly, we fix $d_f = r_{\rm{in}}\sqrt{3}$, effectively fixing the opening angle on the inner edge to $\alpha_{\rm{min}} = 30^o$ (see Fig.\,\ref{fig:geom_definition}). In the geometry considered here this also sets an upper limit on the covering fraction of $f_{\rm{cov}, \rm{max}} \simeq 0.86$. From Fig.\,\ref{fig:lin_profiels} we see that the velocity parameters $r_v$ and $\beta$, as well as the radial launch efficiency parameter $\kappa$, have only a small impact on the overall line shape. Hence we fix these at $r_v=200$, $\beta=1$, and $\kappa=-1$\rev{, as typically done for radiative winds \citep[e.g][]{Matzeu22}}. During initial fitting attempts we also noted that the fit was insensitive to the turbulent velocity $v_{\rm{turb}}$, hence we fix this at 0 (where the choice is motivated by the fact that this allows the code to skip an additional convolution step, speeding up the fit). We also fix all abundances to their solar values, using the values from \citet{Anders89}.

Finally, we include a {\sc diskline} \citep{Fabian89} component, to account for the presence of an inner accretion disc. We allow the inner radius to be free, but fix the outer radius to $10^{3}\,R_G$. We also fix the emissivity index to -3, as expected for a flat disc.

Thus, the final {\sc xspec} model is: {\sc tbabs$_{\rm{MW}}$ * (zcutoffpl$_{\rm{scatt}}$ + tbabs$_{\rm{int}}$ * (pcfabs*zcutoffpl$_{\rm{primary}}$ + rdblur*mytorus + zashift*(xwindfe + diskline)))}, where the {\sc zashift} component accounts for the lack of a redshift parameter in both {\sc xwindfe} and {\sc diskline}. In Appendix\,\ref{app:line_prof_comptest} we test the effect of removing individual model components to the Fe-K$\alpha$ complex in order to assess their significance to the overall fit.

\begin{figure*}
    \centering
    \includegraphics[width=\textwidth]{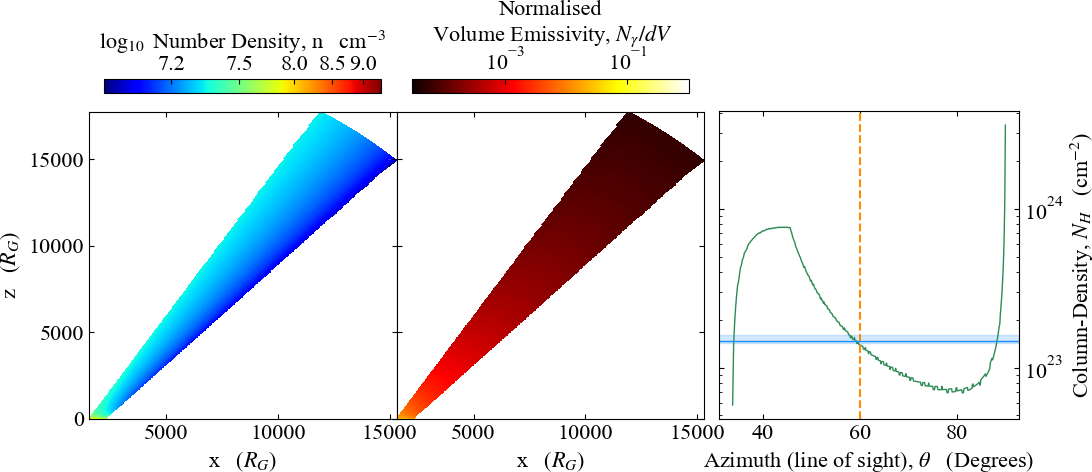}
    \caption{
        \textbf{\textit{Left and Middle:}} Wind density (left) and emissivity (right) profiles corresponding to the best fit {\sc xwindfe} model fit to the \xrism-resolve data (Fig.\,\ref{fig:resolve_fit}) It is clear that the high density and emissivity regions are concentrated towards the base of the wind. However, some additional contribution from larger scale regions in the wind contribute to giving the 'narrow-er' core seen in the line profile in Fig.\,\ref{fig:resolve_fit}. \textbf{\textit{Right:}} Column-density profile as a function of aximuth through the wind (green line). The drop at $\sim 50^{o}$ corresponds to the point where sight-lines do not reach the outer radius of the wind, rather cutting through the central regions. The vertical dashed orange line indicates the observer inclination for NGC\,4151, while the solid blue horizontal line indicates the column-density of the {\sc pcfabs} component as inferred from the continuum fit. The agreement here is remarkable given that the {\sc xwindfe} fit did not use any prior constraints on line-of-sight column-density.
    }
    \label{fig:windGoeom_fit}
\end{figure*} 

This gives an excellent fit to the $5.0-6.6$\,keV \xrism-resolve spectrum, with a $\chi^{2}_{\nu} = 2134/2067 = 1.03$. The resulting fit parameters are given in table\,\ref{tab:resolve_fitpars}, and the spectrum is shown in Fig\,\ref{fig:resolve_fit}, where we have elected to only show the 6.0-6.6\,keV range in order to highlight the Fe-K$\alpha$ complex.

The resulting {\sc xwindfe} parameters suggest a slow-moving ($\log_{10} (v_{\infty}/c) \simeq -3.5$) wind with a moderately low mass-outflow rate ($\log_{10} \dot{m}_w \simeq -3.0$). Specifically the mass-outflow rate is important, as this is significantly lower than the accretion rate for NGC\,4151 ($\dot{m} \sim 0.02$, \citealt{Mahmoud20}), indicating that although the impact on the line-profile is significant, this slow wind's impact to the accretion \rev{flow} will likely be minimal. In some sense this is reassuring, given the near ubiquitous presence of a BLR across the AGN population. 

The inferred outflow velocity, $v_{\infty}/c \simeq 10^{-3.5} \simeq 100$\,km s$^{-1}$, however does appear rather low. Especially given that the escape velocity at $r_{\rm{in}} \simeq 1550$ is $v_{\rm{esc}} \sim 7600$\,km s$^{-1}$. \rev{This is likely because our model \emph{only} considers outflowing material in a wind geometry. More realistic BLR geometries are probably multi-component, with a non-outflowing virialized component and potentially also an inflowing component. Indeed, the physically motivated FRADO models \citet{Czerny11} for the BLR give material that first rises in a dusty wind, before failing and falling back towards the disc. As such, we caution the reader that assuming the entire BLR can be modelled with {\sc xwind}, as done here for demonstrative purposes, is likely physically incomplete, and a more realistic picture could be obtained by including some contribution from non-outflowing material. However, this is beyond the scope of this paper, where we instead note that {\sc xwind} does give a physically motivated method for adding in a wind component to the BLR emission. Additionally, \citet{Kraemer06} does suggest that this wind component should be present, as seen through slow moving, $\sim500$\,km\,s$^{-1}$, material seen in absorption.}

%This is a likely due to the assumption in our model that the entire BLR emission can be described with the same out-flowing wind model. Instead, if the wind fails at some point, as would be expected for $v_{\infty} < v_{\rm{esc}}$ and postulated in the FRADO models of \citet{Czerny11}, then there should be some contribution from non-out-flowing material, which could have the effect of reducing the inferred outflow velocity when using our model. Though previous work by \citet{Kraemer06} does show slow moving material with radial velocities of the order 500\,km s$^{-1}$, seen in absorption, with a suggested origin of a disc wind.  

More interesting in terms of the wind itself is the resulting density profile, as seen in Fig.\,\ref{fig:windGoeom_fit}. Initially, one would think that the low outflow rate would give a relatively low density. However, this is counteracted by the low outflow velocity, giving instead a moderate density range of $\sim 10^7 - 10^9$\,cm$^{-3}$. While the high density regions are concentrated towards the base of the wind, the larger scale regions appear to have a relatively uniform profile, with $n \sim 10^{7.2} - 10^{7.5}$\,cm$^{-3}$. This impacts the line-profile, allowing for significant emission from spatial scales well beyond the base, giving the smooth 'narrowing' of the emission line seen in Fig.\,\ref{fig:resolve_fit}. \rev{However, it is uncertain whether all of this `large spatial scale' emission is genuinely present, and not simply a fitting degeneracy due to a lack of a non-outflowing component to the BLR. Nonetheless, we note that the derived density profile gives a line-of-sight column-density of $N_{H}\sim 1.4\times10^{23}$\,cm$^{-2}$, in surprisingly close agreement of the $1.47^{+0.14}_{-0.05}\times10^{23}$\,cm$^{-2}$ derived for the partial covering absorber in the continuum fit with \xrism-Xtend and NuSTAR. This adds confidence that the derived model geometry is physically consistent with the observed X-ray absorption.}
%It is clear that while the maximum width is set by the inner launch radius, the shape of the line core is instead set by the larger scale emission.

%%%%%%%%%%%%%%%%%%%%%%%%%%%%%%%%%%%%%%%%%%%%%%%%%%%%%%%%%%%%%%
\section{Discussion and Conclusions}
\label{sec:discussion_conclusions}

We have developed a model for the emission line profiles originating from a wind structure at BLR like scales, {\sc xwind}\rev{, which we have made publicly available}. Our approach gives a \rev{physically motivated} model that can easily be fit to \xrism\ data, and in the future NewAthena.

%There are, of course, a number of caveats. Perhaps the most important being that we assume neutral material. This is unlikely to be the case for material surrounding an AGN, given the highly ionising SED typically associated with the inner accretion flow. However, including this would likely require a more sophisticated radiative transfer treatment, which then adds unwanted complexity to the model code, \rev{which is} designed to be fast and simple. The main effects we anticipate from this is some additional smoothing, due to small differences in rest frame energy as the material becomes more/less ionised, as well as small changes in the total fluorescence due to changes in the location of the Fe-K absorption edge.
There are a number of caveats. Perhaps the most important being that we assume neutral material. This is unlikely to be the case for material surrounding an AGN, given the highly ionising SED typically associated with the inner accretion flow. \rev{Given the wide range in wind density, one might expect a range of ionisation states throughout. This should affect the line profile, as even small changes in ionisation can shift or weaken the emitted line \citep[e.g.][]{Bianchi05}, which would in the best case simply mildly smooth the line, in the worst case significantly alter the line shape. The magnitude of these effects should depend on wind properties, placing some limits to model parameters within which the results can be trusted. While we do not calculate ionisation on the fly, we can make a rough estimate. Using the SED of NGC\,4151 from \citet{Mahmoud20} and the wind densities derived in Section\,4.4 we arrive at ionisation parameters ranging from $\log_{10} \xi \sim 2.5 - 3.5$ in the high emissivity regions, where $\xi = L_{\rm{ion}}/(n_H R^2)$ and $L_{\rm{ion}}$ is the integrated luminosity from 13.6\,eV to 13.6\,keV (1--1000\,Ryd); following standard definitions \citep{Ferland17, Mehdipour16b}. This is initially problematic, given that in this regime one expects a significant contribution from ionised species of iron \citep{Mehdipour16b}. But this is neglecting clumping, which could increase the local density of the emitting material, and thus reduce the overall ionisation state. \citet{XRISM25_PDS456} showed conclusively a clumpy outflow in PDS\,456 through absorption, although that is for a UFO wind where the models of \citet{Matzeu22} and \citet{Luminari24} are likely more appropriate. However, it would not be unreasonable to assume that clumping can extend to these slower BLR scale winds. Especially given the presence of the low ionisation slow moving absorption lines seen in \citet{Kraemer06}. As such we caution the reader that this model is a first order approximation. Both in terms of ignoring internal clumping, such that {\sc xwind} can be thought of as a volume-averaged approximation, and in terms of the ionisation state of the material which is completely ignored. While increasing the local density through clumping can address \emph{some} of the caveats, we caution that this model \emph{should not} be used for material significantly inwards of the BLR ($r \lesssim 1000$) where ionisation will almost certainly alter the line shape.}

There is also the caveat of the simplified geometry and velocity profile. We have assumed that the material is always outflowing. However this is \rev{likely} not the case, especially given the rather low outflow velocity derived in section\,\ref{sec:application_xrism}. It may be expected that at some point the outflow fails and the material falls back down, as in e.g. the failed radiative outflow models of the BLR \citep{Czerny11, Naddaf21}. The motion of this infalling material is likely to be somewhat more chaotic than the smooth outflowing wind assumed here, potentially providing some additional smoothing or narrow component to the model line profile. Of course, the total effect of this would depend somewhat of the column-density of the intervening wind, as the wind inwards of this would still cast a shadow across any material falling back down. \rev{Nontheless, we caution that a physically complete picture should include at least a non outflowing component, and that {\sc xwind} only provides a method for adding in the wind.}

%Finally, we have assumed a perfectly smooth wind, which is almost certainly not the case. This is especially clear in \citet{XRISM25_PDS456} which showed conclusively a clumpy outflow in PDS\,456 through absorption; although that is for a UFO wind where the models of \citet{Matzeu22} and \citet{Luminari24} are likely more appropriate than the one presented here. 
%\rev{However, emission should be less sensitive to clumping than absorption, seeing as it integrates over the entire volume, whereas absorption only integrates across a single sight-line. As such our smooth emission only model can be considered a volume-averaged first order approximation.} In future work we will aim to address this more quantitatively. 
%That being said, since we are only modelling the emission line, we suspect that the effects of clumping will be small, if not negligible. 
%This comes from considering the fact that in emission we integrate over the entire wind volume, whereas in absorption one only integrates across a single sight-line which is naturally more sensitive to local inhomogeneities. Nonetheless, in future work we will aim to address this more quantitatively. 

Caveats aside, the model \rev{does give a statistically good fit to} the data, while providing physical constraints on BLR like winds. It is particularly promising that our model fits not only give launch radii consistent with that seen in previous \xrism\ analysis of NGC\,4151 \citep{XRISM24_NGC4151, Xiang25}, but also provides reasonable mass-outflow rates compared to the mass-accretion rate for NGC\,4151. \rev{Though we stress that this is only one viable explanation, and that there are likely other models that give equally good fits to these data}.

\xrism\ has provided an unprecedented view of the X-ray emitting regions in astrophysical sources, and for AGN has allowed for the characterisation of multiple components to the Fe-K$\alpha$ emitting regions. These data are of sufficient quality that they warrant physical models, in order to start characterising their actual structure. Our approach provides a step forward in this regard, giving a fast and simple, yet physical, model for BLR scale outflows. In the future we will start applying our model to a wider sample of \xrism\ observations of moderately accreting AGN, with the aim of understanding the structure of BLR like outflows across the wider population.

{\renewcommand{\arraystretch}{1.5} %Adjust table spacing
\begin{table}
    \centering
    \begin{tabular}{c c c} 
   \textbf{Component} & \textbf{Parameter (Unit)} & \textbf{Value} \\ 
   \hline 
   \hline 
   {\sc rdblur} & Index & -2 \\ 
   & $r_{\rm{in}}$ ($R_G$) & $13074^{+26760}_{-2178}$ \\ 
    & $r_{\rm{out}}$ ($R_{G}$) & $10^{7}$ \\ 
\hline 
   {\sc mytorus} & $N_H$ $10^{24}$\,cm$^{-2}$ & $9.1^{+0.5}_{-6.4}$ \\ 
   & $\Gamma^{\dagger}$ & 1.7103 \\ 
   & Norm. & $0.08^{+0.01}_{-0.02}$ \\ 
\hline 
   {\sc xwindfe} & $\log_{10} \dot{m}_{w}$ & $-3.0^{+0.8}_{-1.3}$ \\ 
   & $r_{\rm{in}}$ ($R_G$) & $1549^{+837}_{-258}$ \\ 
   & $r_{\rm{out}}$ ($R_{G}$) & $=1.5 r_{\rm{in}}$ \\ 
   & $d_{f}$ ($R_{G}$) & $= r_{\rm{in}} \sqrt{3}$ \\ 
   & $f_{\rm{cov}}$ ($\frac{\Omega}{4 \pi}$) & $0.83^{+0.01}_{-0.05}$ \\ 
   & $\log_{10} v_{\infty}$ ($c$) & $-3.5^{+0.7}_{-1.4}$ \\ 
   & $r_v$ ($R_G$) & 200 \\ 
   & $\beta$ & 1 \\ 
   & $v_{\rm{turb}}$ (km/s) & 0 \\ 
   & $\kappa$ & -1 \\ 
   & $A_{\rm{Fe}}$ ([Fe]/[Fe$_{\odot}$]) & 1 \\ 
   & $N_{0}^{*}$ & 0.0766176 \\ 
   & $\Gamma^{\dagger}$ & 1.7103 \\ 
\hline 
   {\sc diskline} & $E_{0}$ (keV) & 6.4 \\ 
   & Index & -3 \\ 
   &  $r_{\rm{in}}$ ($R_G$) & $25.5^{+12.4}_{-8.3}$ \\ 
   & $r_{\rm{out}}$ ($R_{G}$) & $10^{3}$ \\ 
   & Norm. &  $0.00050^{+9e-05}_{-7e-05}$ \\ 
\hline 
   & Inc. (deg) & 60 \\ 
   & Redshift & 0.003326 \\ 
   \hline 
   \hline 
   $\chi^2$/d.o.f & & 2133.8/2067 = 1.03 \\ 
   \hline 
\end{tabular} 

    \caption{Best fit parameters for the model fit to the Fe-K$\alpha$ complex in the \xrism-Resolve spectrum. The continuum parameters are not shown here, since they are previously given in table\,\ref{tab:continuumFit}. The errors correspond to 90\,\% confidence intervals, calculated from an MCMC chain run for 160\,000 steps. \\
    $^{\dagger}$ fixed to $\Gamma_{\rm{primary}}$ in table\,\ref{tab:continuumFit}. \\
    $^{*}$ fixed to Norm.$_{\rm{primary}}$ in table\,\ref{tab:continuumFit}. \\
    }
    \label{tab:resolve_fitpars}
\end{table}
}

%%%%%%%%%%%%%%%%%%%%%%%%%%%%%%%%%%%%%%%%%%%%%%%%%%%%%%%%%%%%%%
\begin{acknowledgements}
    \rev{We thank the anonymous referee for their constructive comments, which improved the quality of this manuscript.}
      We would like to thank the \xrism\ team for their work on the initial compilation and analysis of the high resolution data, which motivated this work. We would also like to thank Jon M. Miller for useful discussions, in particular on the NGC\,4151 data. SH acknowledges support from an IFPU fellowship, as well as previous support from the Science and Technologies Facilities Council (STFC) through the studentship ST/W507428/1. CD acknowledges support from STFC through the grant ST/T000244/1.
        \newline
      This work made use of the following software: {\sc xspec} \citep{Arnaud96}, {\sc numpy} \citep{Harris20}, {\sc astropy} \citep{Astropy13, Astropy18, Astropy22}, and \href{https://mesonbuild.com}{meson}.
\end{acknowledgements}

\section*{Code availability}
The {\sc xwind} model code is available via the corresponding author's GitHub page (\url{https://github.com/scotthgn}). We release two version. One written exclusively in {\sc fortran} designed to be used with the {\sc xspec} spectral fitting package (\url{https://github.com/scotthgn/XWIND}), and a {\sc python} version designed to give more flexibility (and plotting options), referred to as {\sc pyxwind} (\url{https://github.com/scotthgn/PyXWIND}).

\section*{Data availability}
Both the \xrism\ and NuSTAR data are publicly available via the \href{https://heasarc.gsfc.nasa.gov/cgi-bin/W3Browse/w3browse.pl}{HEASARC} service, and were previously published in \citet{Xiang25}.

% Please note that we have included the references below in
% order to compile the document, but we ask you to:

\bibliographystyle{aa}
\bibliography{Refs}

%%%%%%%%%%%%%%%%%%%%%%%%%%%%%%%%%%%%%%%%%%%%%%%%%%%%%%%%%%%%%%%
% Appendices must be placed after   \end{thebibliography}
% They will be placed automatically on a new page.
%%%%%%%%%%%%%%%%%%%%%%%%%%%%%%%%%%%%%%%%%%%%%%%%%%%%%%%%%%%%%%%
\begin{appendix}
%%%%%%%%%%%%%%%%%%%%%%%%%%%%%%%%%%%%%%%%%%%%%%%%%%%%%%%%%%%%%%%
% In the PDF output, floats should be placed
% under their own appendix, not before the title, nor after the
% title of the next appendix.

% In short appendices, onecolumn floats (\figure*
% or \table*) will generate a blank page.
% To prevent this behaviour, a few examples are provided here. 

% In case you have a lot of floating objects for little text and the 
% LaTeX engine moves the floats away from their context, the command
% \FloatBarrier of the “placeins” package will empty the
% float buffer and place all stored floats in the continuity.

% If you still encounter problems with wide floats placement,
% just use the onecolumn environment throughout the appendices.
%%%%%%%%%%%%%%%%%%%%%%%%%%%%%%%%%%%%%%%%%%%%%%%%%%%%%%%%%%%%%%%

\section{{\sc xwind} model documentation}

We release both an {\sc xspec} version of {\sc xwind}, written in {\sc fortran}, as well as a stand-alone {\sc python} version, {\sc pyxwind}, for use-cases outside of {\sc xspec}. Within {\sc xwind} we release three model flavours: {\sc xwindline}, {\sc xwindfe}, and {\sc xwindconv}. Here we give a brief overview of each model flavour. We note that {\sc pyxwind} includes additional functionality for plotting and the calculation of column-density profiles. For these details we refer the reader to the documentation on the GitHub page: \url{https://github.com/scotthgn/PyXWIND}.

{\sc xwindline}: This is the base model, which simply focuses on the line-shape as given by the wind. As such the rest-frame energy is treated as a free parameter, and the rest-frame emission is considered a simple delta-function. Here the normalisation should also be treated as a free parameter, as the model does not consider the output equivalent width. The input parameters for {\sc xwindline} are given in table\,\ref{tab:xwindline_pars}.

{\sc xwindfe}: This model is specific to the Fe-K$\alpha$ complex. As such it calculates self-consistently the equivalent width of the emission line (see section\,\ref{sec:xwind_EW}), and so the normalisation in {\sc xspec} should \emph{always} be fixed at unity. The absolute line flux is instead set by the number of photons absorbed, and re-emitted, by the wind, which is fundamentally governed by the density profile as well as the incident X-ray spectrum originating from the central corona. Hence, the model takes as input parameters the normalisation and photon index of the incident spectrum. \emph{These should always be tied to a corresponding fit of the broad-band continuum, and not left free while fitting the line-profile}. In addition to the self-consistent normalisation, {\sc xwindfe} uses the 7-Lorentzian \citet{Holzer97} profile for the rest-frame emission. This leads to it naturally including Lorentzian wings, as well as both Fe-K$\alpha_1$ and Fe-K$\alpha_2$ lines. The input parameters for this model are given in tabel\,\ref{tab:xwindfe_pars}.

{\sc xwindconv}: Convolution model. Takes the line-profiles from {\sc xwindline}, and uses them as a convolution kernel that is the applied to some input spectrum. Note that the normalisation used is such that the convolution conserves the photon number of the input spectrum (i.e if you convolve an input spectrum containing 100 photons with {\sc xwindconv} the output spectrum will also contain 100 photons). The advantage here is that one can apply wind broadening to either continuum emission (e.g the diffuse continuum, under certain assumptions) or output line emission from e.g more realistic radiative transfer calculations (from e.g {\sc cloudy} \citealt{Ferland17}), again under the assumption that the emissivity follows the profile calculated here. The input parameters are given in table\,\ref{tab:xwindconv_pars}. 

{\renewcommand{\arraystretch}{1.6}
\begin{table}[h]
    \centering
    \begin{tabular}{c p{4cm} c}
         Parameter & Description & Units \\
         \hline
         $\log_{10} \dot{m}_w$ & Wind mass outflow rate & $\dot{M}_w/\dot{M}_{\rm{Edd}}$ \\
         $r_{\rm{in}}$ & Inner launch radius (see Fig.\,\ref{fig:geom_definition}) & $R_{G}$ \\
         $r_{\rm{out}}$ & Outer launch radius & $R_{G}$ \\
         $d_f$ & Distance from origin to wind focus (see Fig.\,\ref{fig:geom_definition}) & $R_{G}$ \\
         $f_{\rm{cov}}$ & Wind covering fraction as seen from the central black hole & $\frac{\Omega}{4\pi}$ \\
         $\log_{10} v_{\infty}$ & Outflow velocity at infinity (see Eqn.\,\ref{eqn:vl}) & $c$ \\
         $r_{v}$ & Velocity scale length. The distance along the streamline where the wind reaches half of $v_{\infty}$ & $R_{G}$ \\
         $\beta$ & Wind velocity exponent. Determines the acceleration of the wind (see Eqn.\,\ref{eqn:vl}) & Dimensionless \\
         $v_{\rm{turb}}$ & Turbulent velocity. Assumed constant across the wind. Sets the width of the Gaussian smoothing kernel used to emulate turbulence. & km/s \\
         $\kappa$ & Sets the weighting for the wind launch efficiency as a function of radius. Larger values will weigh the mass-outflow towards larger radii (and vice-versa for lower values). See Eqn.\,\ref{eqn:dMdot_dA} & Dimensionless \\
         Inc. & Observer inclination, measured from the z-axis & degrees \\
         E0 & Rest-frame energy of the emission line & keV \\
         Norm. & Normalisation. Sets the total number of photons within the line & photons s$^{-1}$ cm$^{-2}$ \\
         \hline
    \end{tabular}
    \caption{Parameters in {\sc xwindline}. These are given in the order they should be passed to the code.}
    \label{tab:xwindline_pars}
\end{table}
}

{\renewcommand{\arraystretch}{1.6}
\begin{table}[h!]
    \centering
    \begin{tabular}{c p{4cm} c}
         Parameter & Description & Units \\
         \hline
         $\log_{10} \dot{m}_w$ & Wind mass outflow rate & $\dot{M}_w/\dot{M}_{\rm{Edd}}$ \\
         $r_{\rm{in}}$ & Inner launch radius (see Fig.\,\ref{fig:geom_definition}) & $R_{G}$ \\
         $r_{\rm{out}}$ & Outer launch radius & $R_{G}$ \\
         $d_f$ & Distance from origin to wind focus (see Fig.\,\ref{fig:geom_definition}) & $R_{G}$ \\
         $f_{\rm{cov}}$ & Wind covering fraction as seen from the central black hole & $\frac{\Omega}{4\pi}$ \\
         $\log_{10} v_{\infty}$ & Outflow velocity at infinity (see Eqn.\,\ref{eqn:vl}) & $c$ \\
         $r_{v}$ & Velocity scale length. The distance along the streamline where the wind reaches half of $v_{\infty}$ & $R_{G}$ \\
         $\beta$ & Wind velocity exponent. Determines the acceleration of the wind (see Eqn.\,\ref{eqn:vl}) & Dimensionless \\
         $v_{\rm{turb}}$ & Turbulent velocity. Assumed constant across the wind. Sets the width of the Gaussian smoothing kernel used to emulate turbulence. & km/s \\
         $\kappa$ & Sets the weighting for the wind launch efficiency as a function of radius. Larger values will weigh the mass-outflow towards larger radii (and vice-versa for lower values). See Eqn.\,\ref{eqn:dMdot_dA} & Dimensionless \\
         Inc. & Observer inclination, measured from the z-axis & degrees \\
         $A_{\rm{Fe}}$ & Iron abundance, relative to solar. Uses the abundance values from \citep{Anders89} & [Fe]/[Fe$_{\odot}$] \\
         $N_0$ & Normalisation of the incident X-ray power-law emission & photons s$^{-1}$ cm$^{-2}$ at 1\,keV \\
         $\Gamma$ & Photon index of the incident X-ray power-law spectrum & Dimensionless \\
         \hline
    \end{tabular}
    \caption{Parameters in {\sc xwindfe}.}
    \label{tab:xwindfe_pars}
\end{table}
}

{\renewcommand{\arraystretch}{1.6}
\begin{table}[h!]
    \centering
    \begin{tabular}{c p{4cm} c}
         Parameter & Description & Units \\
         \hline
         $\log_{10} \dot{m}_w$ & Wind mass outflow rate & $\dot{M}_w/\dot{M}_{\rm{Edd}}$ \\
         $r_{\rm{in}}$ & Inner launch radius (see Fig.\,\ref{fig:geom_definition}) & $R_{G}$ \\
         $r_{\rm{out}}$ & Outer launch radius & $R_{G}$ \\
         $d_f$ & Distance from origin to wind focus (see Fig.\,\ref{fig:geom_definition}) & $R_{G}$ \\
         $f_{\rm{cov}}$ & Wind covering fraction as seen from the central black hole & $\frac{\Omega}{4\pi}$ \\
         $\log_{10} v_{\infty}$ & Outflow velocity at infinity (see Eqn.\,\ref{eqn:vl}) & $c$ \\
         $r_{v}$ & Velocity scale length. The distance along the streamline where the wind reaches half of $v_{\infty}$ & $R_{G}$ \\
         $\beta$ & Wind velocity exponent. Determines the acceleration of the wind (see Eqn.\,\ref{eqn:vl}) & Dimensionless \\
         $v_{\rm{turb}}$ & Turbulent velocity. Assumed constant across the wind. Sets the width of the Gaussian smoothing kernel used to emulate turbulence. & km/s \\
         $\kappa$ & Sets the weighting for the wind launch efficiency as a function of radius. Larger values will weigh the mass-outflow towards larger radii (and vice-versa for lower values). See Eqn.\,\ref{eqn:dMdot_dA} & Dimensionless \\
         Inc. & Observer inclination, measured from the z-axis & degrees \\
         \hline
    \end{tabular}
    \caption{Parameters in {\sc xwindconv}.}
    \label{tab:xwindconv_pars}
\end{table}
}

\section{Comparison to {\sc diskline}}
\label{app:diskline_comp}

\begin{figure*}[h]
    \centering
    \includegraphics[width=\textwidth]{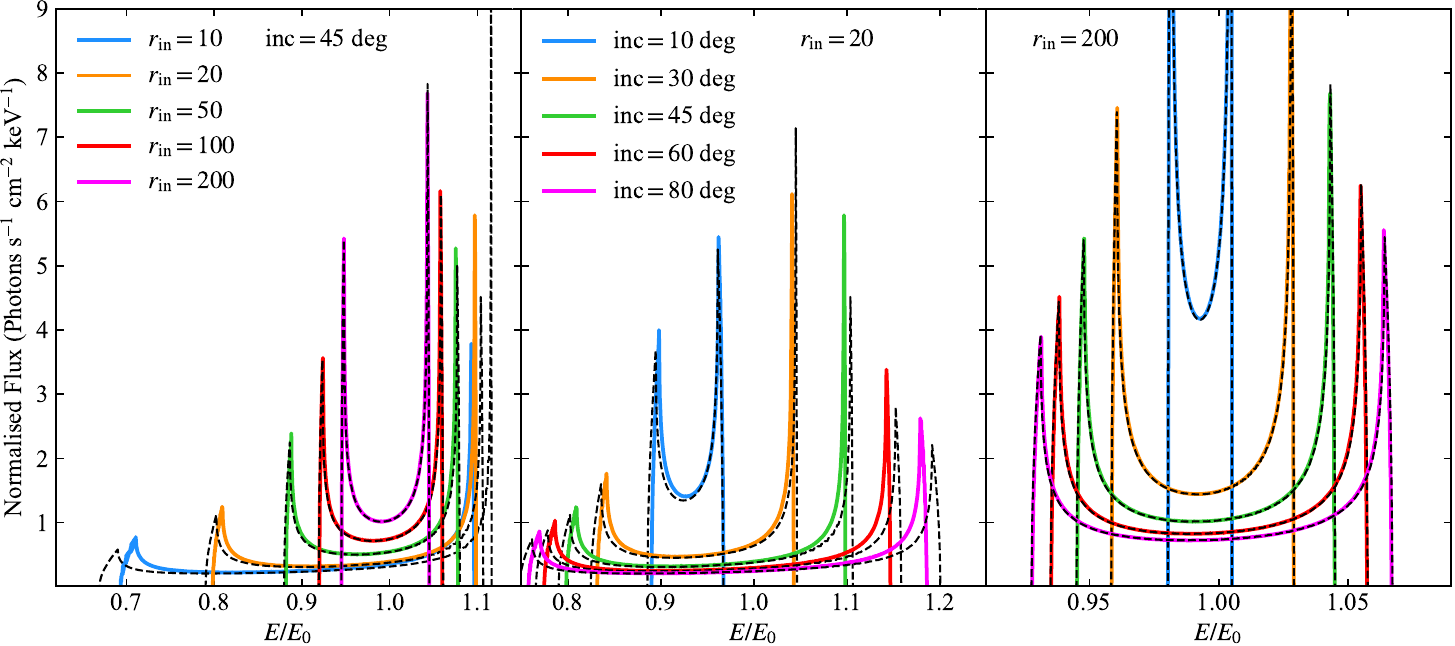}
    \caption{
    Comparison of line profiles for a Keplerian disc, calculated considering only relativistic Doppler shifts as outlined in section\,\ref{sec:line_prof} (solid coloured lines) and calculated using the {\sc diskline} model \citep{Fabian89} (dashed lines), which includes some corrections expected from the Schwarzchild metric. In all cases these are calculated for a narrow annulus, such that $r_{\rm{out}} = 1.1r_{\rm{in}}$. \emph{The left} panel shows the profiles while stepping through values of $r_{\rm{in}}$, in order to highlight where our approximations break down. For $r_{\rm{in}} \gtrsim 50$ our line profile matches that or {\sc diskline} well, whereas below this we start to see systematic offsets originating from our choice of metric. \emph{The middle and right} panels show the profiles while stepping through observed inclination, with $r_{\rm{in}}=20$ and $200$ respectively. Again, in the weak field limit, where general relativistic corrections are small, our approximation matches well that of {\sc diskline}.
    }
    \label{fig:disklineComp}
\end{figure*}

As a sanity check we, we adapt our model to a Keplerian disc, allowing for a direct comparison to the well established {\sc diskline} \citep{Fabian89}. The main differences from {\sc xwind} are that we now only consider the Keplerian velocity component in Eqn.\,\ref{eqn:Efrac} (as well as the gravitational redshift). We also calculate the emission using a simple radial emissivity profile, with $\epsilon(r) \propto r^{-3}$. 

The resulting line-profiles are shown in Fig.\,\ref{fig:disklineComp}, where we have performed comparisons while stepping through inner radius (left panel), and observer inclination (middle and right panels). {\sc diskline} is shown as the dashed black lines, while the profile from our code is shown as the solid coloured lines. For large radii, $r \gtrsim 50$, the profiles are entirely consistent with one another. This is not the case at small radii where we start to see some systematic deviation. This is likely due to out choice of a flat-space time metric. While {\sc diskline} ignores light-bending, it does still calculate the energy shifts and impact parameters in the Schwarzchild metric. This becomes important at small radii. As such, we do not recommend the use of our model for very small launch radii (as might be expected from e.g. at UFO type wind).

Nonetheless, at BLR scales, the general relativistic corrections are negligible, with perhaps the exception of the gravitational redshift, which is included in our model.

\section{Line-profile component tests}
\label{app:line_prof_comptest}

\begin{figure*}
    \centering
    \includegraphics[width=0.85\textwidth]{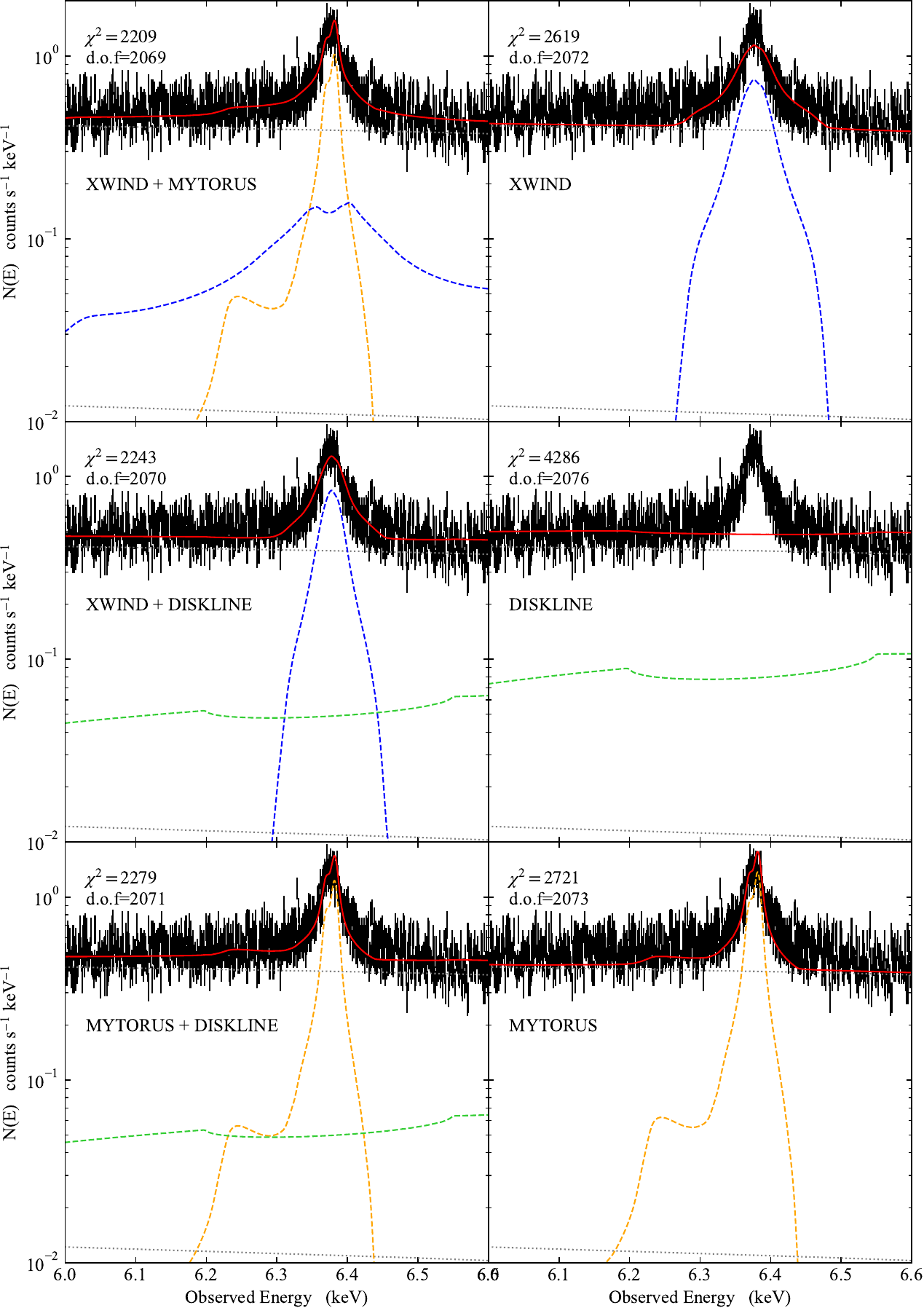}
    \caption{Model fits to \xrism-resolve for subsets of our total line-profile model, to highlight the significance of each component. As with the main model in Fig.\,\ref{fig:resolve_fit} these have been fit over the 5.0-6.6\,keV energy range. The left column shows the case for a two-component model to the Fe-K$\alpha$ complex, while the right shows a single component scenario. The relevant model components (shown as dashed coloured lines) are {\sc xwindfe} (blue), {\sc mytorus} (orange), and {\sc diskline} (green). In all cases, these result in a statistically significantly worse fit than our fiducial three component model for the Fe-K$\alpha$ complex.}
    \label{fig:resolve_modelComp_spec}
\end{figure*}

\begin{figure*}
    \centering
    \includegraphics[width=0.85\textwidth]{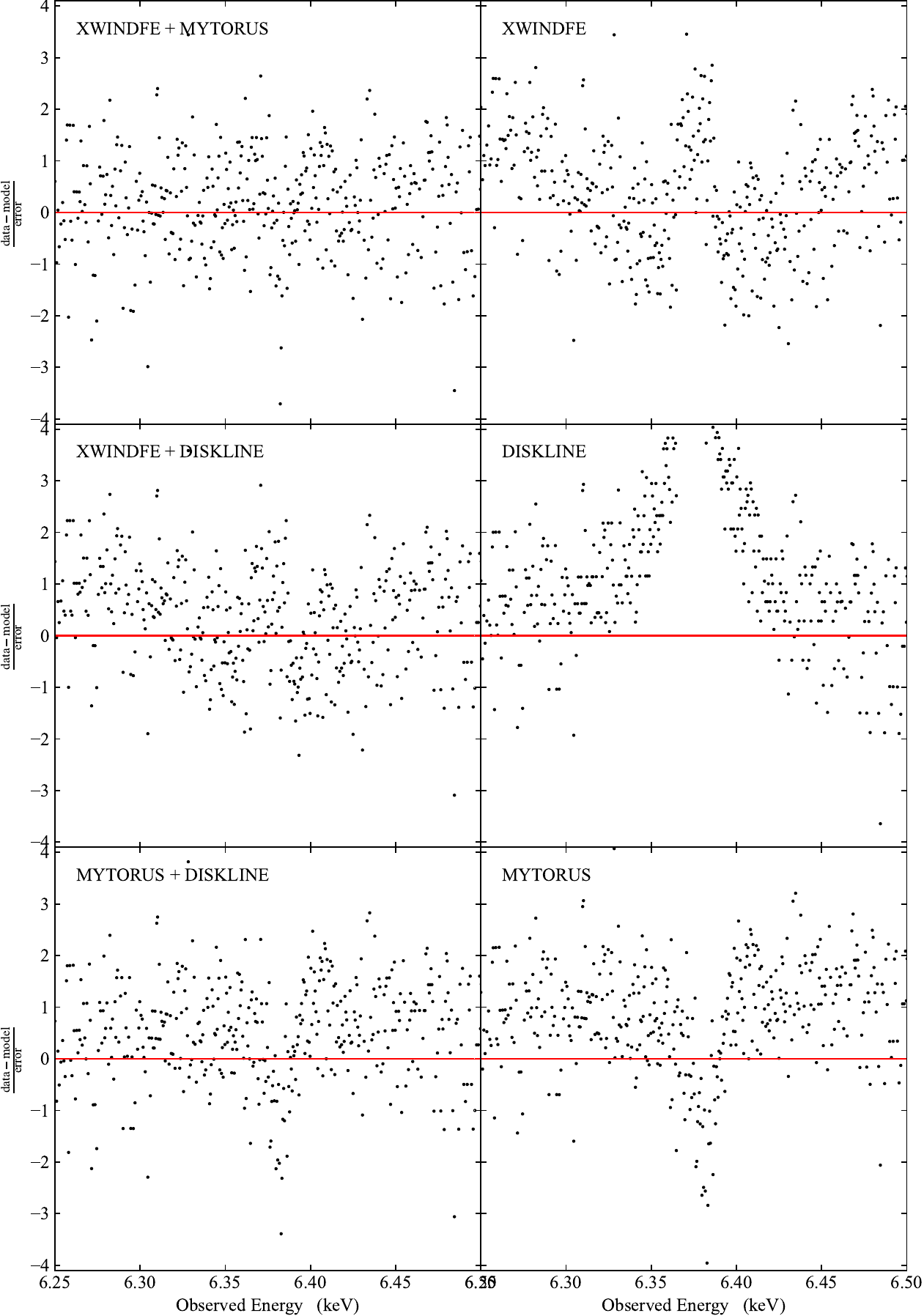}
    \caption{Residual to the fits presented in Fig.\,\ref{fig:resolve_modelComp_spec}. Note we have chosen to zoom in on the 6.25-6.5\,keV region in order to highlight the residuals around the Fe-K$\alpha$ complex. The fits, however, are performed across the 5.0-6.6\,keV range.}
    \label{fig:resolve_modelComp_residual}
\end{figure*}

We present here a brief set of comparison fits to quantitate the significance of each component. In total we perform six additional fits, removing/including the {\sc xwindfe}, {\sc mytours}, and {\sc diskline} components. The underlying continuum model is fixed to that presented in section\,\ref{sec:application_xrism}. As with our fiducial model fit, we use the $5.0-6.6$\,keV range of the \xrism-resolve spectrum. The resulting fits are shown in Fig.\,\ref{fig:resolve_modelComp_spec}, with a a zoom in on the residual shown in Fig.\,\ref{fig:resolve_modelComp_residual}. Each panel in the figures gives the relevant model combination to the line profile.

Perhaps the least significant component to the line profile is {\sc diskline}, with $\Delta \chi^2 =75$ with only 2 additional degrees of freedom. This is not particularly surprising, given that broad components suffer from dilution with the underlying continuum, especially in high resolution spectra.

More importantly for this paper, we see both the {\sc mytorus} and {\sc xwindfe} components are significant, with $\Delta \chi^2 = 109$ and $\Delta \chi^2 = 145$ for 3 and 4 additional degrees of freedom respectively. It is quite clear in residuals shown in Fig.\,\ref{fig:resolve_modelComp_residual} that these components are required to model the narrow and intermediary components to the line.

\end{appendix}
\end{document}